\documentclass[]{aastex631}

\usepackage[T1]{fontenc}

\usepackage{multirow}
\usepackage{amsmath}
\usepackage{hhline}
\usepackage{array, makecell}
\usepackage{booktabs, tabularx}
\usepackage{comment}
\usepackage{hyperref}
\usepackage{multirow}

\usepackage{xcolor}

\newcommand{\gray}{$\gamma$-ray}
\newcommand{\grays}{$\gamma$-rays}
\newcommand{\mmdc}{{\texttt{MMDC}}}

\usepackage{xcolor}
\newcounter{tr}
\ifnum \value{tr}>5

\fi

\begin{document}

\title{Markarian Multiwavelength Data Center (MMDC): A Tool for Retrieving and Modeling Multi-temporal, Multi-wavelength and Multi-messenger Data from Blazar Observations}

\author[0000-0003-2011-2731]{N. Sahakyan}
\affiliation{ICRANet-Armenia, Marshall Baghramian Avenue 24a, Yerevan 0019, Armenia}

\author[0000-0002-3777-7580]{V. Vardanyan}
\affiliation{ICRANet-Armenia, Marshall Baghramian Avenue 24a, Yerevan 0019, Armenia}

\author[0000-0002-2265-5003]{P. Giommi}
\affiliation{Associated to INAF, Osservatorio Astronomico di Brera, via Brera, 28, I-20121 Milano, Italy}
\affiliation{Center for Astrophysics and Space Science (CASS), New York University Abu Dhabi, PO Box 129188 Abu Dhabi, United Arab Emirates}
\affiliation{Institute for Advanced Study, Technische Universit{\"a}t M{\"u}nchen, Lichtenbergstrasse 2a, D-85748 Garching bei M\"unchen, Germany}

\author[0000-0003-4477-1846]{D. B\'egu\'e}
\affiliation{Bar Ilan University, Ramat Gan, Israel}

\author[0000-0002-5804-6605]{D. Israyelyan}
\affiliation{ICRANet-Armenia, Marshall Baghramian Avenue 24a, Yerevan 0019, Armenia}

\author[0000-0002-5804-6605]{G. Harutyunyan}
\affiliation{ICRANet-Armenia, Marshall Baghramian Avenue 24a, Yerevan 0019, Armenia}

\author[0009-0007-4567-7647]{M. Manvelyan}
\affiliation{ICRANet-Armenia, Marshall Baghramian Avenue 24a, Yerevan 0019, Armenia}

\author[0009-0007-7798-2072]{M. Khachatryan}
\affiliation{ICRANet-Armenia, Marshall Baghramian Avenue 24a, Yerevan 0019, Armenia}

\author[0000-0002-8852-7530]{H. Dereli-B\'egu\'e}
\affiliation{Bar Ilan University, Ramat Gan, Israel}

\author[0000-0002-0031-7759]{S. Gasparyan}
\affiliation{ICRANet-Armenia, Marshall Baghramian Avenue 24a, Yerevan 0019, Armenia}

\begin{abstract}
The Markarian Multiwavelength Data Center (\mmdc) is a web-based tool designed for accessing and retrieving multiwavelength and multimessenger data from blazar observations. \mmdc\ facilitates the construction and interactive visualization of time-resolved multi-band spectral energy distributions (SEDs) of blazars by integrating: \textit{(i)} archival data from over 80 catalogs and databases, \textit{(ii)} optical data from all-sky survey facilities such as ASAS-SN, ZTF, and Pan-STARRS, and \textit{(iii)} newly analyzed datasets in the optical/UV band from \textit{Swift}-UVOT, in the X-ray band from \textit{Swift}-XRT and NuSTAR observations, and the high-energy $\gamma$-ray band from \textit{Fermi}-LAT observations. \mmdc\ distinguishes itself from other online platforms by the large quantity of available data. For instance, it includes data from all blazar observations by \textit{Swift} and NuSTAR, as well as the results of detailed spectral analysis in the $\gamma$-ray band during different emission states, covering the period from 2008 to 2023. Another important distinguishing feature of \mmdc\ is its ability to enable precise, self-consistent theoretical modeling of the observed data using machine learning algorithms trained on leptonic and lepto-hadronic models, which consider the injection of particles and all relevant cooling processes. \mmdc\ is an innovative tool which significantly enhances blazar research by providing a comprehensive framework for data accessibility, analysis, and theoretical interpretation, thereby advancing our understanding of blazar emissions and the underlying astrophysical processes.
\end{abstract}



\section{Introduction}

As in all disciplines, be it of
science, humanities, history or others, data plays a crucial
role in research: it is the 
foundational basis for all discoveries.
Specifically in the field of astrophysics, data is commonly used for measuring the positions, velocities, and compositions
of celestial objects \citep[e.g.][]{2016A&A...595A...1G, 2023A&A...674A...1G}, for
investigating the evolution and structure of local stars, distant galaxies and of the Universe \citep{ 2020A&A...641A...1P}, as well as for testing the
theoretical predictions of different models. Data in
astrophysics not only takes the form of observations, but can also
be the results of advanced numerical simulations \citep[e.g.][]{2022ApJ...930L..16E, 2019ApJ...875L...5E, 2019ComAC...6....2N}, catalogues of model
parameters obtained from detailed model fitting \citep[e.g.][]{ BGB19, 2020ApJ...893...46V, 2022ApJS..263...24A} , etc. The accurate
analysis and interpretation of the data are essential for advancing our
understanding of the cosmos. This also implies that data should
be straightforwardly available to any scientists who wish to test
a newly developed model or interpret new observations.

Astrophysical observations, collected through a multitude of different observatories,
enable researchers to develop and validate models
of various celestial phenomena. In recent years, the deployment of
numerous high-precision instruments, relying on
always more sophisticated observational techniques, has significantly
increased the volume and the acquisition rate of astrophysical data. The advancement of both space- and
ground-based observatories, the development of large-scale survey projects,
as well as automated sky survey programs such as the Sloan Digital Sky Survey
\citep{2000AJ....120.1579Y} or the upcoming Vera C. Rubin Observatory\footnote{\url{https://rubinobservatory.org}} are generating vast
amounts of data, leading to an exponential increase of their volume and enabling detailed studies of
the universe on an unprecedented scale. These technological advancements
have not only increased the quantity of acquired astrophysical data
but have also enhanced its quality, thereby facilitating more comprehensive
and more accurate analyses of cosmic phenomena.

Despite this significant increase in both
the quality and quantity of astrophysical data, there are still severe
challenges faced by the scientific community in extracting
the maximum of information contained in the observations. Indeed, data is accumulated
by many different instruments, telescopes and
observatories, each with its own unique particularities and method
of data analysis. Maximizing the insight into the data requires the integration
and analysis of broad methods, which necessitates a deep understanding of
the various data types and the methodologies appropriate for their analysis.
For example, understanding the multiwavelength properties of a single
active galactic nucleus (AGN) requires detailed analysis
of data accumulated in the radio, optical, X-ray, and \gray\ bands. Moreover, the
large volumes and complexity of the data sets often require advanced
computational tools for efficient and timely analysis. Therefore, effectively
extracting the information contained in astrophysical
data requires not only technological advancements but also a robust framework
for data management and analysis.

Astrophysical data is often accessible to the scientific community through
online databases, such as the NASA/IPAC Extragalactic Database\footnote{\url{https://ned.ipac.caltech.edu}}, SSDC SED Builder\footnote{\url{https://tools.ssdc.asi.it/SED/}}, Firmamento \citep{2024AJ....167..116T}, etc. Over time, they have evolved significantly, transforming from resources accessible to a limited
number of users to being widely available resources for the research community.
Often, the data were accessible only to
researchers involved in the collaborations responsible for using the
telescopes. However, the adoption of open-access policies has led to a
substantial increase in availability and consequently in users, which
has, in turn, increased the potential for scientific discovery.

Furthermore, the parallel advancements in Machine Learning
(ML) algorithms are revolutionizing the analysis of observed
data. They can efficiently handle large and
complex data sets to uncover patterns and correlations that
traditional methods might miss, and perform efficient and on the fly-source
classification. For example, in \citet{2023MNRAS.519.3000S}, blazar candidates of uncertain type were classified using different ML models (Artificial Neural Networks, XGBOOST, and LIGHTGBM algorithms) trained on both the spectral and temporal properties of  already classified blazar classes. ML can also be used for the interpretation and analysis
of data via the creation of surrogate models of expensive numerical models, allowing
parameter exploration and constraints of these state-of-the art models \citep[e.g.][]{BvL23, 2024ApJ...963...71B, 2024arXiv240207495S, 2024A&A...683A.185T}.
Successful integration of these ML methods into online databases
promises to enhance the ability to interpret vast amounts
of heterogeneous data, leading to new
astrophysical discoveries and insights into how the cosmos works.

This paper introduces the Markarian Multiwavelength DataCenter (\mmdc), a new
online tool developed to facilitate the accessibility and modeling of
multiwavelength data for blazar research. \mmdc\ allows users (i)
to build time-resolved multiwavelength spectral energy distributions (SEDs) of blazars, using data from multiwavelength catalogues as well as new data analyzed in optical/UV, X-ray, and High Energy (HE; $>100$ MeV) \gray\ bands, (ii) to
visualize these SEDs interactively, and (iii) to perform theoretical modeling of the SEDs. This theoretical modeling relies on a recently proposed ML framework for blazars SED analysis \citep{2024ApJ...963...71B, 2024arXiv240207495S} and enables in-depth modeling of the observed data by self-consistent models. Incorporating a large amount of newly analyzed
data and providing the possibility of theoretical modeling using Neural Networks (NNs), \mmdc\ can significantly contribute to advance blazar research.

The paper is structured as follows: Section \ref{blazar} presents a multiwavelength view of the blazars. Section \ref{data} presents the archival, optical/UV, and X-ray data available through \mmdc. Section \ref{gdata} presents the \gray\ data available through \mmdc. Section \ref{mmdc} introduces \mmdc\ and its tools, describing its structure and main components.  The multiwavelength and multitemporal SEDs retrieved from \mmdc\ are detailed in Section \ref{mseds}, while Section \ref{modeling} introduces the modeling of the SEDs using a neural network. The conclusions and future perspectives are provided in Section \ref{conc}.

\section{A multiwavelength view of blazars}
\label{blazar}

Among the various types of AGNs, blazars stand out due
to their powerful emissions, which originate from relativistic jets
oriented at small viewing angles relative to the observer \citep{1995PASP..107..803U}.
This alignment causes the emission from blazar jets to be strongly Doppler
amplified towards the observer, enabling
the detection of blazars even at high redshifts \citep[e.g., see][]{2017ApJ...837L...5A,
2020MNRAS.498.2594S, 2023MNRAS.521.1013S, 2024MNRAS.528.5990S}. The emission of blazar is highly variable across most bands \citep[e.g.,][]{2013ApJ...762...92A, 2014Sci...346.1080A,
2016ApJ...824L..20A, 2018ApJ...854L..26S}, underlying the complex processes behind the jet dynamics as well as behind the 
particle acceleration and emission. For instance, recent temporal
analyses of several sources have revealed periodic variability in the
\gray\ band \citep[e.g.,][]{2020ApJ...896..134P, 2023A&A...672A..86R}.

\begin{figure}
    \centering
    \includegraphics[width=0.5\linewidth]{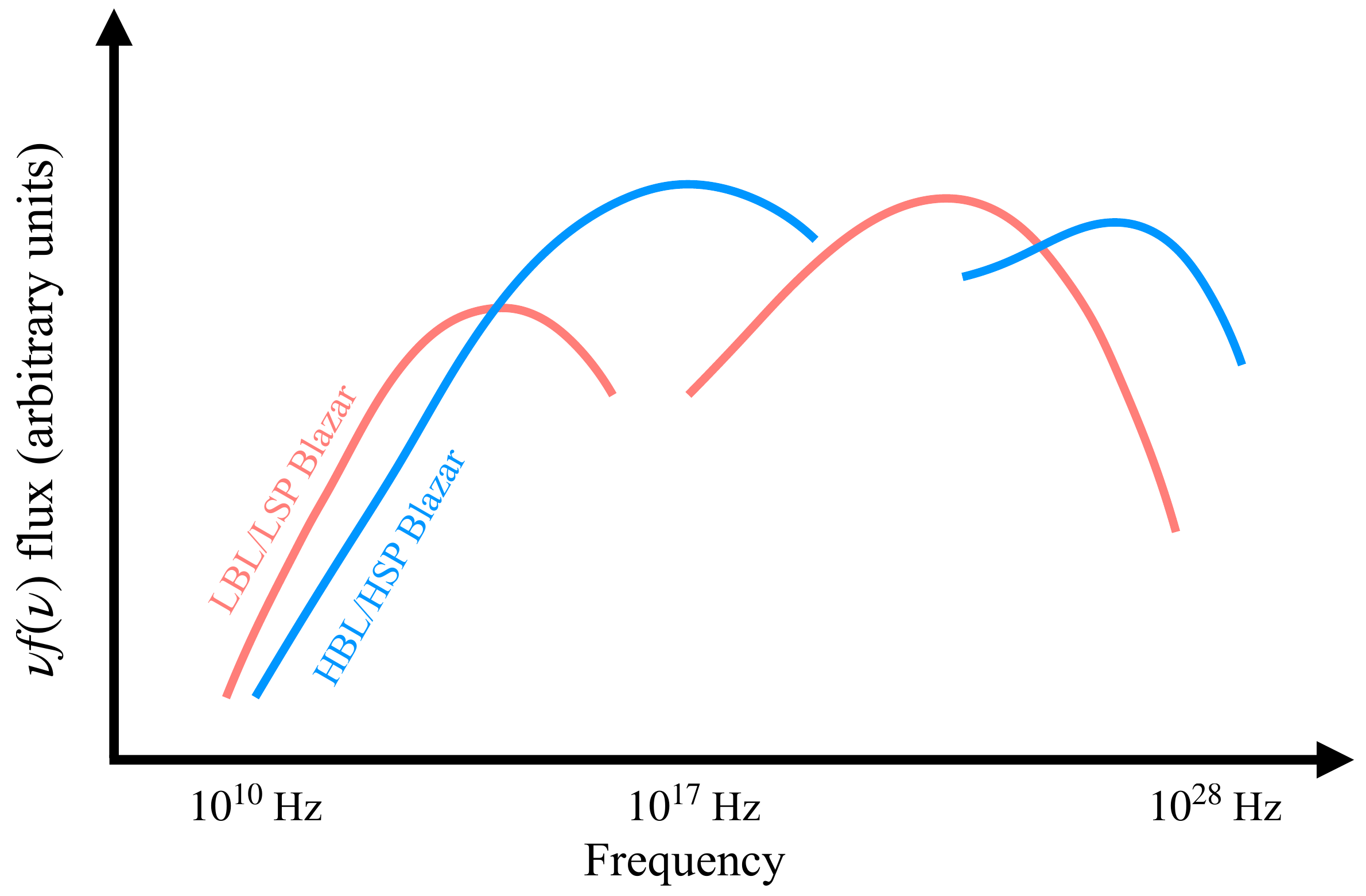}
    \caption{The SED of different types of blazars, showing that the emission extends from radio to VHE \gray\ bands. The SED of LSP or LBL blazars is shown in red, and that of HSP or HBL blazars is shown in light blue. The figure is adapted from \citet{2017A&ARv..25....2P}.}
    \label{fig:blazar}
\end{figure}

Blazars, being powerful and bright sources,
are frequently monitored by various instruments, yielding
a wealth of observations. While data from individual bands
offer insights into the specific process which
emission dominates in that limited energy range, combining data
from all accessible bands provides a comprehensive view of
all relevant processes occurring within the relativistic
jet. A blazar SED typically
extends from the radio to the very
high energy (VHE; $>100$ GeV) \gray\ bands, exhibiting a double-peaked
morphology \citep[see Fig. \ref{fig:blazar} and][]{2017A&ARv..25....2P}. The first
low-energy peak, usually between the far infrared and X-rays,
is believed to result from synchrotron emission of non-thermal
electrons within the jet, while the second component, peaking between
X-rays and VHE \grays, could either originate from the interaction of these
non-thermals electrons with local synchrotron or external photons, \citep{1985A&A...146..204G, 1992ApJ...397L...5M,
1996ApJ...461..657B, 1994ApJ...421..153S, 1992A&A...256L..27D,
1994ApJS...90..945D, 2000ApJ...545..107B} or result from
hadronic processes requiring jets to be loaded with protons \citep[e.g.,][]{1993A&A...269...67M, 1989A&A...221..211M, 2001APh....15..121M, mucke2, 2013ApJ...768...54B, 2015MNRAS.447...36P, 2022MNRAS.509.2102G}.

The peak frequency of the first component $\nu_p$ is used to classify blazars into
low, intermediate, and high-energy synchrotron peaked sources
\citep{Padovani1995, Abdo_2010, 2021Univ....7..492G}. These classes are characterized by $\nu_p<10^{14}$Hz for low synchrotron peaked sources (LSPs or LBLs), $10^{14} < \nu_p < 10^{15}$Hz for intermediate synchrotron peaked sources (ISPs or IBLs) and $\nu_p > 10^{15}$Hz  for high synchrotron peaked sources (HSPs or HBLs). Fig. \ref{fig:blazar} is a plot showing the SED of LSP/LBL blazars (in red) and HSP/HBL blazars (in light blue), highlighting the difference in the peak of their synchrotron component. Historically, blazars have also been grouped based on their optical lines as Flat Spectrum Radio Quasars (FSRQs) when the optical lines are strong, and as BL Lacertae type objects (BL Lacs) when the lines are faint or absent.
The accumulated multi-temporal and multi-wavelength observational datasets are
crucial for identifying new blazar candidates and characterizing their temporal
and spectral properties of the emissions across different bands. The data
across various bands enables the investigation of emission variability
and the determination of correlations or anti-correlations between these different bands.
This is crucial for understanding particle interactions and emission within the jet.
Modeling the observed data with time-dependent leptonic
or hadronic models offers insights into the composition,
structure and dynamics of the relativistic jets,
allowing the inference of the physical conditions
and processes occurring within them \citep[see e.g.,][]{2008ApJ...686..181F,2013A&A...558A..47C, 2014ApJ...782...82D, 2020A&A...640A.132M, 2021MNRAS.504.5074S,
2022MNRAS.517.2757S, 2022MNRAS.513.4645S, 2023A&A...670A..49M, 2024A&A...681A.119R}. The advent of multimessenger
astrophysics — specifically neutrinos for blazar
observations — and the accessibility of science-ready, multi-temporal, and
multi-wavelength data alongside self-consistent modeling tools, open a new
window for in-depth investigation of the physics of blazar jets.

\section{Data in \mmdc}\label{data}

As discussed in the Introduction, the extensive data now available from blazar observations has the potential to greatly enhance our understanding of the emission processes within these sources. In this section, we describe the archival, optical/UV and X-ray data available in \mmdc.
\subsection{Archival data}

Blazars have been frequently observed
in different bands, either during monitoring programs or through individual
source observations. As a result, a large volume of data is available
through various catalogues. To access the archival data, the {\it VOU-Blazars}
V2.22 is used. It is a significant evolution of the original
{\it VOU-Blazars} tool \citep{2020A&C....3000350C}, specifically developed
for the Firmamento \citep[\url{http://firmamento.hosting.nyu.edu},][]
{2024AJ....167..116T} and \mmdc\, platforms, together with different
python scripts which access and retrieve additional data.
This tool is designed to identify blazars based on the SED features that differentiate blazars from other astronomical sources. Then, the tool builds the SEDs of the identified blazars using
public multi-wavelength
photometric and spectral data accessible through the International
Virtual Observatory Alliance (IVOA) services. For a given position in the sky,
the tool performs a cone search around the position in different catalogues
(from radio to HE \gray\ bands), retrieves all available data,
converts it to common SED units and corrects for absorption in the Galaxy.
For the most recent catalogues queried via the {\it VOU-Blazars} tool,
see Table 1 in \citet{2024AJ....167..116T}. The large number of
catalogues accessible from {\it VOU-Blazars} ensures that the
archival data requested are as complete as possible
and provide comprehensive information on blazar emission properties
across a broad range of frequencies.

\subsection{Optical and UV data}\label{opt}

In the optical/UV band, several facilities collect data from blazar,
including the All-Sky Automated Survey for Supernovae
\citep[ASAS-SN;][]{2014ApJ...788...48S, 2017PASP..129j4502K}, the Zwicky
Transient Facility \citep[ZTF;][]{2019PASP..131a8002B}, the Panoramic
Survey Telescope and Rapid Response System \citep[Pan-STARRS1;][]
{2016arXiv161205560C} and the Neil Gehrels Swift Observatory \citep{2004ApJ...611.1005G},
(hereafter \textit{Swift}) with its Ultra-violet Optical Telescope (UVOT). Data from ASAS-SN, ZTF, and
Pan-STARRS1 are obtained at search time
when the source name is entered into the search box. In contrast, data
from \textit{Swift}-UVOT observations underwent a thorough analysis before being included
in the database of \mmdc{}, see below.
\begin{itemize}
  \item {\it ASAS-SN:} the survey ASAS-SN, comprising various stations across both
  hemispheres, is designed to observe
  the entire visible sky nightly to a depth of V$\sim17$ mag. Until late
  2018, observations were conducted in the $V$ band, subsequently switching
  to the $g$ band. The primary objective of ASAS-SN is to detect bright
  transients throughout the visible sky; however, the data accumulated
  during the survey also facilitate studies of blazar
  emissions. Following each observation, the image processing pipeline,
  utilizing the ISIS image-subtraction package \citep{1998ApJ...503..325A}, 
  is used to perform the photometry on all targets, providing the community with near-instantaneous
  data within one hour post-observation. Data from the archive is retrieved
  using the Python client Sky Patrol V2.0 \citep{2023arXiv230403791H}. All data within a \(5^{\prime\prime}\) radius of the target position are downloaded and further corrected for Galactic extinction
  in accordance with \citet{1999PASP..111...63F}
  \citep[see also ][]{2020A&C....3000350C} \footnote{A similar script, \texttt{mag2flux}, is available at \url{https://github.com/ecylchang/VOU_Blazars.}}. Finally the magnitudes are
  converted into fluxes. This process provides access to blazar observations
  from 2011 together with the corresponding appropriate references. This dataset is then
  aggregated with other data provided by \mmdc, and made available for download and further analysis.

  \item ZTF scans the northern sky at high cadence (approximately 2 days)
  in the \(g\), \(r\), and \(i\) bands using a \(47 \, \mathrm{deg}^2\)
  wide-field imager mounted on a 48-inch Schmidt telescope on Mount Palomar.
  For a given sky position, data are downloaded via an API
  request\footnote{\url{https://irsa.ipac.caltech.edu/docs/program_interface/ztf_api.html}}
  from the ZTF archive \citep{2019PASP..131a8003M}. This process performs
  a query search around the specified position and extracts data in the
  \(g\), \(r\), and \(i\) bands for objects within a \(5^{\prime\prime}\)
  radius of the target position. Once downloaded
  by \mmdc, the magnitudes are then converted into fluxes for each
  filter, and extinction corrections are
  made applying the extinction rule described
  in \citet{1999PASP..111...63F}, following the method also outlined
  in \citet{2020A&C....3000350C}. The processed data is plotted in \mmdc\
  along with data from all other bands and is accessible for download
  (with corresponding references).
 
  \item {\it Pan-STARRS1:} Pan-STARRS1 is a wide-field imaging facility
  designed to survey the sky for transient and variable phenomena. Located
  near the summit of Haleakala on the Island of Maui, Pan-STARRS uses a
  1.8-meter telescope equipped with the world's largest digital camera,
  comprising almost 1.4 billion pixels. Typically, each night, Pan-STARRS1 covers
  about 1,000 square degrees of the night sky. This is achieved through a
  sequence of four exposures spanning approximately one hour, utilizing
  five filters ($g$, $r$, $i$, $z$ and $y$). The second data release (DR2)
  includes a detection catalog that encompasses all multi-epoch
  observations. Data for the source is downloaded via an API 
  request\footnote{\url{https://catalogs.mast.stsci.edu/docs/panstarrs.html}},
  performing searches for objects within a \(5^{\prime\prime}\) radius
  of the target position. The downloaded dataset is then corrected for
  extinction using the method outlined in \citet{2020A&C....3000350C}.
  This enables the retrieval of data from blazar observations in the
  frequency range of $(3.12-6.23)\times10^{14}$ Hz, which are displayed
  alongside other data in \mmdc. 
  
 \item {\textit{Swift}-UVOT:} The \textit{Swift} telescope, with its exceptional capability
 of monitoring the sky in both the optical/UV and X-ray bands, is a crucial
 instrument for identifying and constraining emission components
 in blazar SEDs. The \textit{Swift}-UVOT telescope can produce images in the V
 (500-600 nm), B (380-500 nm), U (300-400 nm), W1 (220-400 nm),
 M2 (200-280 nm), and W2 (180–260 nm) filters. We have processed all
 blazars observations performed by \textit{Swift}
 between December 2004 and March 2024, amounting to 3009 sources for which UVOT
 observations are available. The list of blazars observed by \textit{Swift} (used also for the
 XRT analysis) was determined by cross-matching the
 master list of blazars \citep{2019A&A...631A.116G} — all blazars from the
 BZCAT 5th edition \citep[5BZCAT;][]{2015Ap&SS.357...75M}
 and all those observed
 in the \gray\ band by \textit{Fermi}-LAT - with the public archive of \textit{Swift}. For
 these identified sources, data were downloaded and reduced using
 an automated script that performs a traditional UVOT analysis\footnote{
 The analysis follows all steps described at \url{https://www.swift.ac.uk/analysis/uvot/}}.
 Specifically, for each observation, photometry was
 computed by selecting counts from a circular region of $5''$ around the
 source, while background counts were estimated from a $20''$ region away
 from the source. Each single image was visually inspected to ensure that
 the source and background selection were not affected by other sources,
 etc. The magnitudes were derived using the {\texttt{uvotsource}} tool,
 and the fluxes were obtained using conversion factors provided by
 \citet{2008MNRAS.383..627P}, and further corrected
 for extinction using the reddening coefficient $E(B-V)$ from the Infrared
 Science Archive\footnote{\url{http://irsa.ipac.caltech.edu/applications/DUST/}}.
 As a result, 19897 ObsIDs were processed,
 which allowed us to estimate 100335 fluxes
 in the optical and UV filters.
\end{itemize}

\subsection{X-ray data}\label{xray}

In \mmdc, in addition to archival X-ray data, new data from \textit{Swift}-XRT (covering the energy range of 0.3-10 keV) and NuSTAR (covering the energy range of 3-79 keV) observations of blazars are provided.

\begin{itemize}
  \item In \citet{2021MNRAS.507.5690G}, the \textit{Swift}-XRT data for all blazars
  observed by \textit{Swift} at least 50 times between December 2004 and the end of
  2020 were analyzed, and the analysis results for 65 blazars
  were made public\footnote{\url{https://openuniverse.asi.it/blazars/swift}}.
  We have extended this analysis to all blazars observed by \textit{Swift} from the beginning of the mission up to March 2024, and we reduced all available X-ray data using Swift\_xrtproc tool. This tool automates \textit{Swift}-XRT
  data analysis from raw data handling to produce science-ready data. It
  includes automated data and calibration file downloading, exposure map
  generation, source and background spectral file creation, and photometric
  analysis across multiple energy bands. It also performs pile-up correction,
  when needed, and uses the XSPEC package to perform spectral fitting,
  converting spectral data for SED and flux estimation. For a detailed
  description of the swift\_xrtproc tool and analysis methodologies,
  see \citet{2021MNRAS.507.5690G}. In total, 4163
  blazars were observed by \textit{Swift} at least once with exposure higher than 200 seconds, and we processed a total
  of 17958 \textit{Swift}-XRT observations. The results of this analysis, i.e., the SED points computed assuming a power-law model, are accessible through \mmdc. 
  \item The first X-ray catalogue of blazars observed by NuSTAR was presented
  in \citet{2022MNRAS.514.3179M}. The catalogue includes 253 observations of
  126 blazars (with 30 blazars being observed multiple times), encompassing
  all blazars observed in the hard X-ray band by September 30th, 2021. We
  have extended the analysis to also include blazars that were observed from
  October 2021 to  30 March 2024. This has been accomplished by cross-matching
  the master list of blazars \citep{2019A&A...631A.116G} with the public archive of NuSTAR.
  Over this additional period, namely between 2021 and
  2024, 66 blazars have been observed, for a total of 102 observations. The analysis
  was performed using the NuSTAR\_Spectra pipeline, which is a
  shell-script designed to automate the analysis of NuSTAR data. It
  includes automated data retrieval, generation of calibrated scientific
  products, and spectral analysis using predefined models such as a power law
  and a logarithmic parabola. The script efficiently manages data extraction
  and processing, optimizing the spectral extraction regions based on
  source brightness to ensure the optimum results. It performs
  comprehensive spectral fitting (assuming a power-law model in this case), identifying optimal spectral parameters
  and producing high-quality post-processed data for SED plotting and modeling. For more details on the tool and analysis methods,
  see \citet{2022MNRAS.514.3179M}. These spectral datasets from all blazars
  observed by NuSTAR can be retrieved via \mmdc.
\end{itemize}
\begin{table*}
    \centering
\caption{The first 15 sources selected for our analysis, listing the name, associated 4FGL designation, class, SED type, right ascension (R.A.), declination (Dec.), and redshift for each source.}
\label{tab:sources}
\begin{tabular}{lccccclc}
\toprule
Source name &  4FGL name & Source class &  SED class   & R.A.  & Dec.  & Redshift\\
\hline
3C 454.3 & 4FGL J2253.9+1609 & FSRQ & LSP & 343.5 & 16.15 & 0.86\\
3C 279 & 4FGL J1256.1-0547 & FSRQ & LSP & 194.04 & -5.79 & 0.54\\
CTA 102 & 4FGL J2232.6+1143 & FSRQ & LSP & 338.15 & 11.73 & 1.04\\
Mkn 421 & 4FGL J1104.4+3812 & BLL & HSP & 166.12 & 38.21 & 0.03\\
PKS 1510-089 & 4FGL J1512.8-0906 & FSRQ & LSP & 228.21 & -9.11 & 0.36\\
PKS 1424-41 & 4FGL J1427.9-4206 & FSRQ & LSP & 216.99 & -42.11 & 1.52\\
NGC 1275 & 4FGL J0319.8+4130 & RDG & LSP & 49.96 & 41.51 & 0.02\\
BL Lac & 4FGL J2202.7+4216 & BLL & LSP & 330.69 & 42.28 & 0.07\\
S5 0716+71 & 4FGL J0721.9+7120 & BLL & ISP & 110.49 & 71.34 & 0.13\\
PKS 2155-304 & 4FGL J2158.8-3013 & BLL & HSP & 329.71 & -30.23 & 0.12\\
PKS 0426-380 & 4FGL J0428.6-3756 & BLL & LSP & 67.17 & -37.94 & 1.11\\
PKS 1502+106 & 4FGL J1504.4+1029 & FSRQ & LSP & 226.1 & 10.5 & 1.84\\
PKS 0454-234 & 4FGL J0457.0-2324 & FSRQ & LSP & 74.26 & -23.41 & 1.0\\
PG 1553+113 & 4FGL J1555.7+1111 & BLL & HSP & 238.93 & 11.19 & 0.36\\
4C +21.35 & 4FGL J1224.9+2122 & FSRQ & LSP & 186.23 & 21.38 & 0.43\\
\hline
\end{tabular}
\end{table*}
\section{\textit{Fermi}-LAT $\gamma$-ray data in \mmdc}
\label{gdata}

In the \gray\ band, the entire sky is continuously monitored by the \textit{Fermi}
satellite. Launched in 2008, the satellite is equipped with two instruments:
the Large Area Telescope (LAT) and the Gamma-ray Burst Monitor (GBM). The LAT
is the primary instrument on board and is designed to scan the entire sky in
the \gray\ band, while the GBM is used to study gamma-ray bursts. The LAT
is a pair-conversion \gray\ telescope, sensitive to
energies ranging from 20 MeV to 300 GeV with a field of view of approximately
2.4 steradians. It primarily operates in an all-sky scanning mode,
and as such is an ideal instrument to study the HE properties of
various astronomical sources, including blazars. For further information on
the LAT instrument, see \citet{2009ApJ...697.1071A}.
\begin{figure*}
     \centering
     \includegraphics[ width=0.94\textwidth]{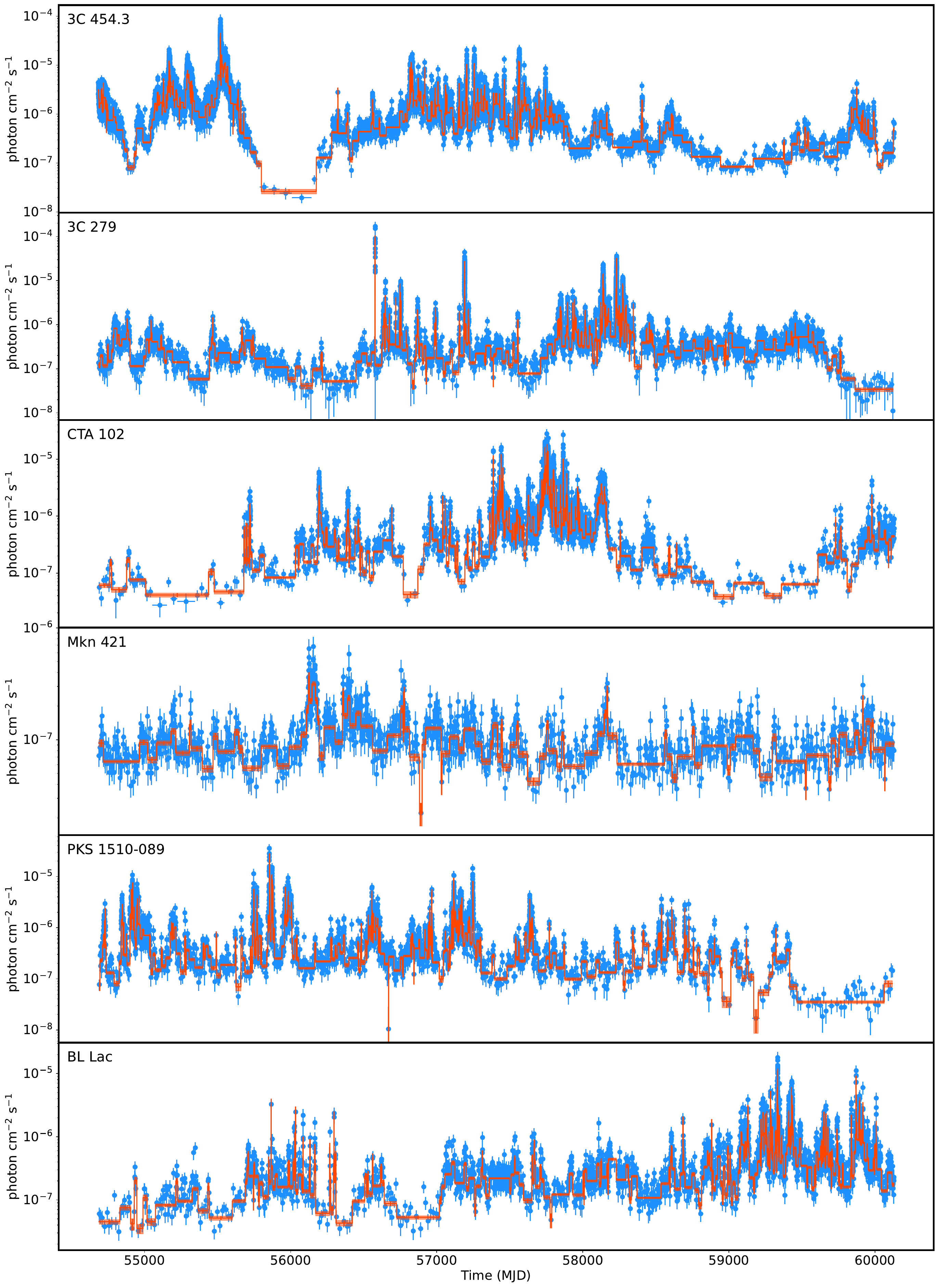}
     \caption{\gray\ light curves of selected sources shown together with their Bayesian blocks binning. This illustrates the large flux variability of blazars as well as the intervals selected for spectral analysis, which follow the blazar temporal variability.}
     \label{gray_LC}
\end{figure*}
The continuous observations by \textit{Fermi}-LAT provide a detailed view of the \gray\ emission from various sources. The recent catalog of \gray\ sources observed by \textit{Fermi}-LAT, namely the fourth
\textit{Fermi}-LAT catalog incremental version \citep[4FGL-DR3, ][]{2022ApJS..260...53A}
contains 6658 sources. The largest population of these sources,
approximately 55.2\%, are blazars of different types. In \mmdc, detailed \gray\ data analysis results are provided for as many sources as possible that exhibit changes in their \gray\ emission states. The analysis initially focused on sources that are bright in the \gray\ band. To achieve this, blazars from the fourth catalog of active galactic nuclei detected by the \textit{Fermi}-LAT \citep[Data Release 3][]{2022ApJS..263...24A} were sorted by decreasing integrated energy flux above 100 MeV, followed by a detailed analysis of the data for each source. However, the analysis was restricted to those blazars for which
the application of Bayesian blocks to the adaptively binned light curves
yielded more than 5 intervals (see Appendix \ref{appendix}). As of the writing of this paper,
466 sources satisfy this last criteria and their data have been analyzed. The analysis is
ongoing and will encompass additional sources. The list of the first
15 brightest sources included in our analysis is provided in Table \ref{tab:sources}.
\begin{figure*}
     \centering
    \includegraphics[width=0.98\textwidth]{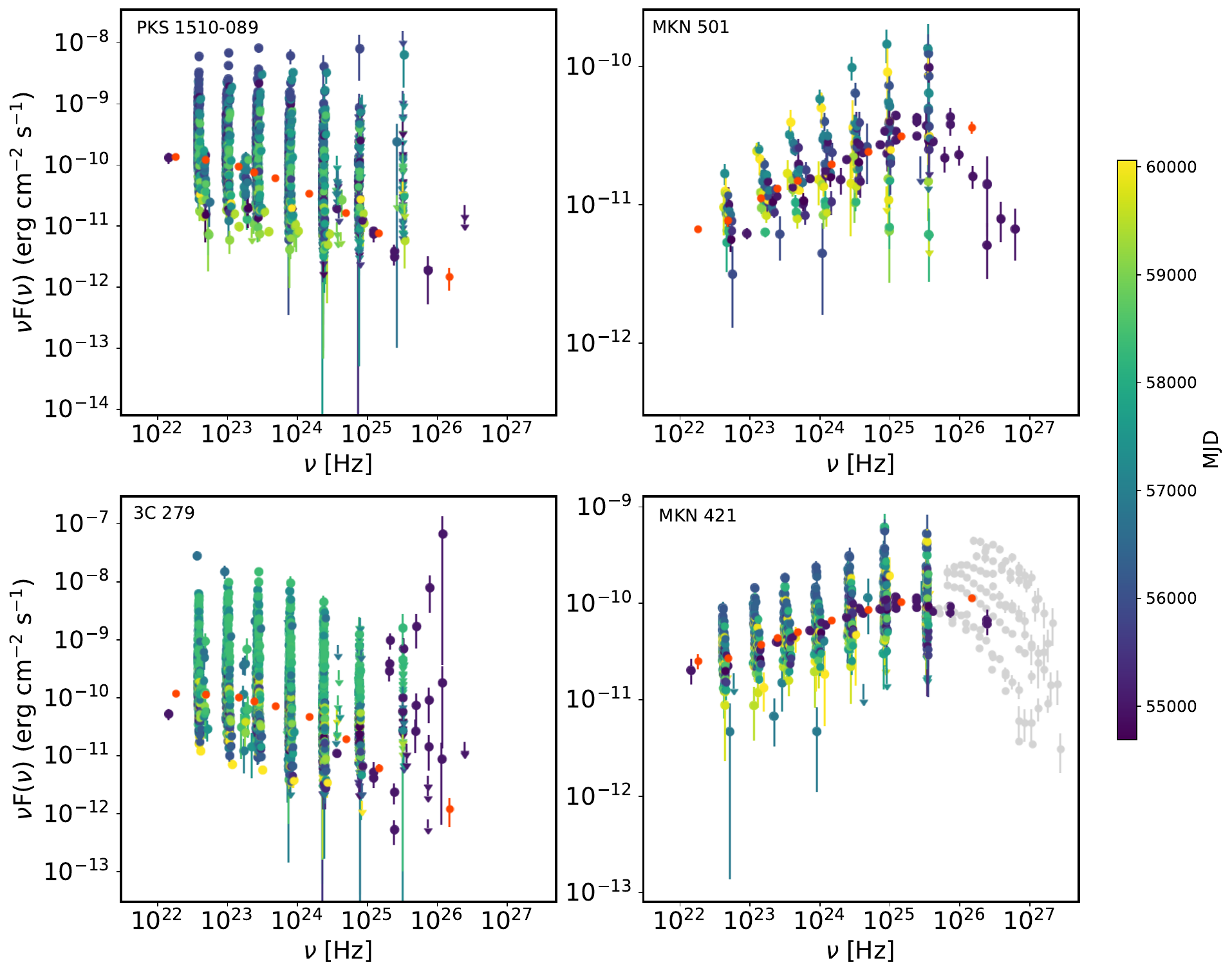}
     \caption{The broadband SEDs of PKS 1510-089 (top left), Mrk 501 (top right), 3C 279 (bottom left), and Mrk 421 (bottom right). The red points represent the averaged \gray\ spectrum retrieved from the 4FGL catalog, while the colored data points are derived from \mmdc, with a different color for each analysed period, represented by the color-bar. Grey data points in the bottom right panel represent the archival TeV data. }
     \label{SED}
\end{figure*}

Currently, the \gray\ data collected between August 4, 2008, and July 4, 2023, are considered for the analysis, with the results available in MMDC. For details on the analysis and applied methods, see Appendix \ref{appendix}. Generating the light curves and performing the spectral
analysis for the considered sources required the
analysis of approximately 128.7k time bins and
96.9k periods for spectral analysis. This analysis is fully automatized and 
ongoing as new observations are made available thereby constantly increasing the number
of time bins, spectral periods and eventually sources. Then, the
produced results are visually inspected and the log
files are checked before the results are uploaded to
the \mmdc\ database.

The adaptively binned light curves for 3C 454.3, 3C 279, CTA 102, Mkn 421,
PKS 1510-089, and BL Lacertae (BL Lac) are shown in Fig. \ref{gray_LC}. These sources
display different emission states in the \gray\ band (flaring and quiescent),
captured by the Bayesian blocks.
For example, the adaptively binned light curve for 3C 454.3 results in 10,872
time intervals and 430 Bayesian intervals. Similarly, for 3C 279 there are
5,460 and 360 intervals, respectively, for CTA 102, there are 4,327 and 353
intervals, respectively, and for MKN 421, there are 1,447 and 103.
As 3C 454.3, 3C 279 and CTA 102 show strong variability, the number of
Bayesian blocks is therefore large, while for Mkn 421,
with moderate flux variation, this number is low. 

During the brightening periods, the photon index also changes (either softening
or hardening), resulting in a spectrum in the \gray\ band that significantly
differs from the time-averaged spectrum presented in the 4FGL catalog. This
spectral variability is evident in Fig. \ref{SED}, which displays the SEDs of
PKS 1510-089 (top left), Mrk 501 (top right), 3C 279 (bottom left), and Mrk
421 (bottom right). The red points represent the time-averaged
data from the 4FGL catalog, i.e., data obtained from the other data centers
(with the grey data in the bottom right panel representing historical TeV
data for Mkn 421). The color-coded data are retrieved from \mmdc, with the
corresponding time periods indicated by the color-bar in the rightmost panel. The spectral
variability in these sources exhibits distinct signatures: the powerful FSRQs,
PKS 1510-089 and 3C 279 (displayed in the top and bottom left
panels in Fig. \ref{SED}), show significant changes in the \gray\ band. The
flux varies from a few times $10^{-12}\:{\rm erg\:cm^{-2}\:s^{-1}}$ to a
few times $10^{-8}\:{\rm erg\:cm^{-2}\:s^{-1}}$, with
both steep and hard spectra. In contrast, Mrk 501 and Mrk 421 (top and bottom
right panels in Fig. \ref{SED}), while characterized
by a hard \gray\ spectrum, do not exhibit a significant flux
variation in that band. However, periods
when the \gray\ spectrum becomes hard are still observable.
The SEDs illustrated in Fig. \ref{SED} highlight the advantages of the time-resolved dataset in the \gray\ band, offering quantitative insights into \gray\ flux changes over long a period. These data product are readily provided by \mmdc.

\section{MMDC: a tool for retrieving, building and modeling time-resolved blazar SEDs} \label{mmdc}

\begin{figure*}
     \centering
    \includegraphics[width=0.98\textwidth]{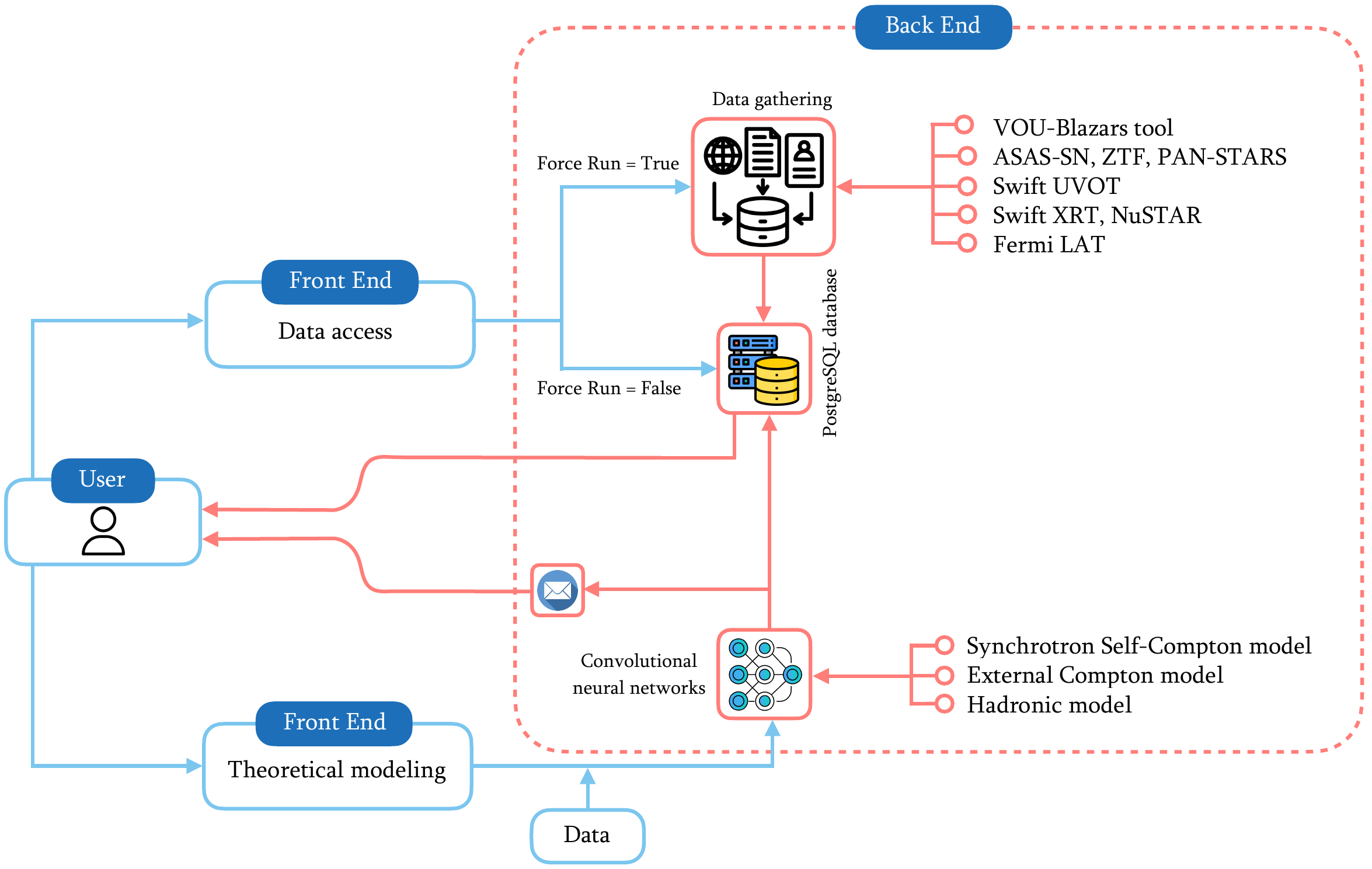}
     \caption{Schematic view of the architecture of \mmdc\ with its main components. The architecture includes the front-end components for data access and theoretical modeling, and the back-end components for data gathering, processing and fitting. The database is managed by PostgreSQL. Through the front-end interface, the user can access data already stored in the PostgreSQL database (if Force Run = False) or initiate a data gathering run (if Force Run = True). The user can also submit data for fitting using CNN, with the results returned through the front-end interface.}
     \label{struct}
\end{figure*}

\mmdc\ is a web-based tool dedicated to retrieving time-resolved multiwavelength
data from blazar observations, allowing to perform in-depth studies of the origin of their emission. 
Fig. \ref{struct} shows the structure of \mmdc\ with its main components.
The two main functionalities are: (i) retrieving time-resolved multiwavelength
data from blazar, resulting from observations across the
entire electromagnetic spectrum, and (ii) modeling the broadband SEDs of
blazars, either from the retrieved SEDs or from user-supplied data. Each
component is described below.

\subsection{The Front-end Web Pages}

The services provided by \mmdc\ are accessible through the dedicated webpage
\url{www.mmdc.am}. The interface of \mmdc, hereinafter referred to as frontend, is designed with a
minimalist approach and prioritizes user-friendliness and efficiency. By
avoiding unnecessary complexity, it ensures that users, regardless of their
technical proficiency, can easily navigate through the various tools and
services provided. Novel tools are implemented to facilitate the possibility
for users to interact directly with the data, enabling them to grasp its
complex content more effectively. The site includes 7 tabs, one for each functionality of \mmdc:

\begin{enumerate}
    \item Home: Introduces the site and provides general information.
    
    \item About: Provides general information about blazars, their emissions, and the specific data available in \mmdc.
    
    \item Data Access: This section enables users to search for and retrieve data of any blazar of interest by providing its name or sky coordinate. %

    \item Theoretical Modeling: This section enables users to model broadband SEDs using Convolutional Neural Networks (CNNs). 
    
   \item Articles: This section allows users to search for articles either by using the name of a source or by performing a cone search around known coordinates.
    
    \item Team: This section presents general information about the individuals who are contributing to the development of \mmdc.
    
    \item Contact: This section enables feedback or communication with the development team.
\end{enumerate}

\subsection{Data Access}

\begin{figure}
    \centering
    \includegraphics[width=0.98\linewidth]{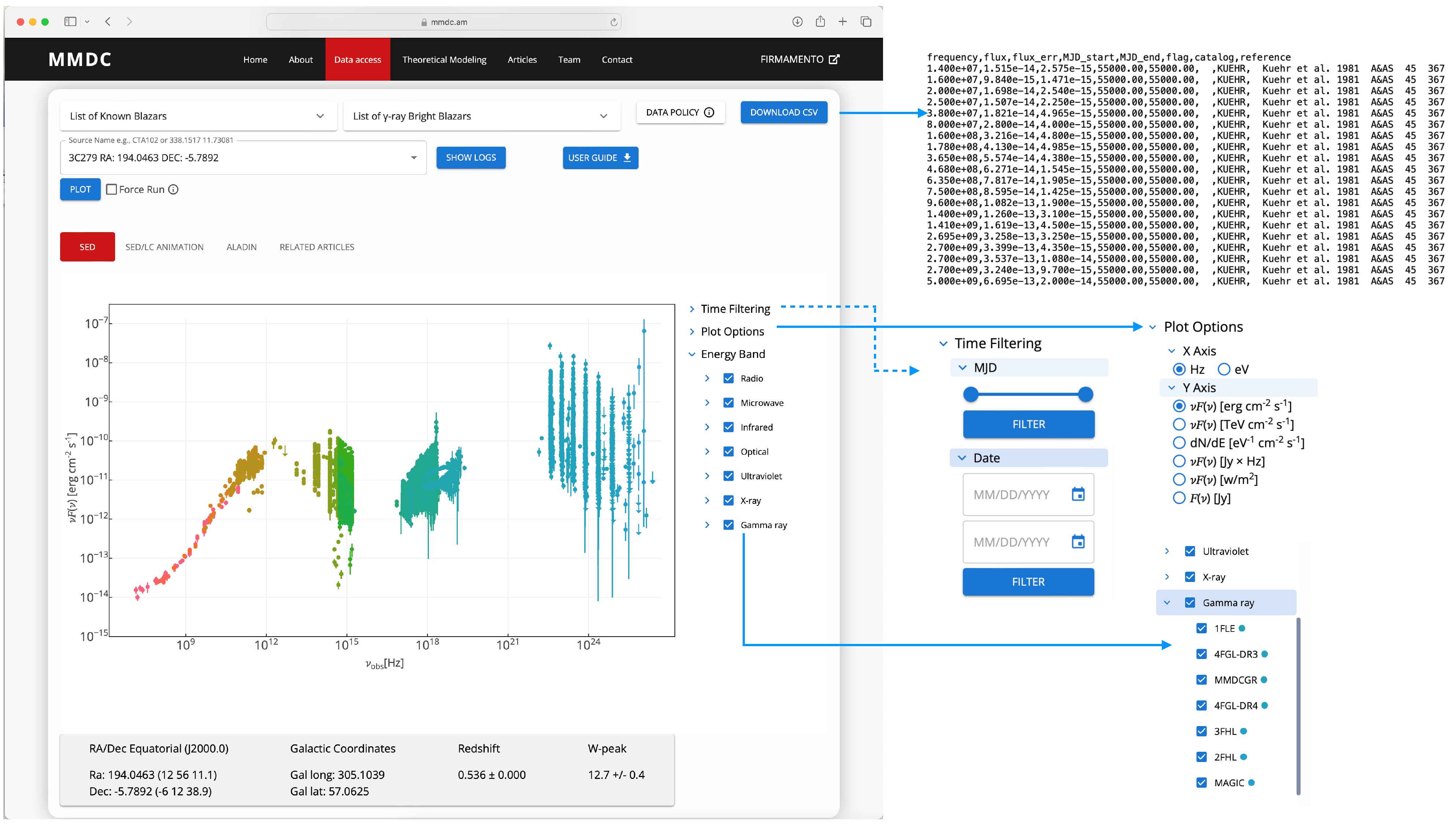}
    \caption{The \mmdc\ interface for data access. The SED can be retrieved by providing the source name or a sky position in the 'Source Name' field. In the top right, an example of part of the data retrieved from \mmdc\ is shown. The interface allows users to filter data based on time ('Time Filtering') and customize the display with various options for axis adjustments ('Plot Options'). Under the energy band section, the catalogs from which the data were retrieved are listed (an example is shown at the bottom right). The bottom panel under the SED provides source information, including RA/Dec coordinates, galactic coordinates, redshift, and the estimated peak frequency of the low-energy synchrotron component \citep[W-peak;][]{2024ApJ...963...48G}.}
    \label{fig:data_acess}
\end{figure}

Fig. \ref{fig:data_acess} shows the Data Access tab, which is used to
retrieve detailed multiwavelength data from blazar observations.
\mmdc\ provides multiwavelength data when the user provides the common name
of a blazar or its sky coordinates in the source name field.
Alternatively, blazars can be selected from the table named
'List of Known Blazars', which includes 6,400 objects from
the 5BZCAT \citep{2015Ap&SS.357...75M}, the 3HSP \citep{2019A&A...632A..77C},
and the \textit{Fermi}-LAT \citep[Data Release 3][]{2022ApJS..263...24A}
catalogues. Blazars can also be chosen from the 'List of \gray\ Bright Blazars',
which contains sources for which \mmdc\ includes SEDs computed over
multiple periods. As the source
name or coordinates is provided, the data is retrieved
from the PostgreSQL database immediately if available (i.e., if the source
SED has already been searched by other users). However, if
'Force Run = True' is selected, the data-gathering process is initiated,
accessing different catalogs and returning the results. This
option is particularly useful if the user wishes to force data retrieval,
especially when the logs indicate that some catalogs were previously
unreachable via the VOU-blazars tool. The real-time software processing
can be monitored by clicking the 'SHOW LOGS' button. Depending on the
amount of available data, this process can take anywhere from a few tens
of seconds to up to 2 minutes.

Once the data are retrieved, the SED is plotted using the Plotly
library\footnote{\url{https://plotly.com}}\, which allows for direct
interaction with the figure. An example of SED generation for 3C 279 is
shown in Fig. \ref{fig:data_acess}, where the data in different bands are
visualized using different color coding. The interactive nature of the data
visualization allows users to explore the SEDs: users can
zoom in and out to examine specific data ranges of interest or exclude
data from any catalog via a  dedicated panel under the energy range
(an example is shown in the bottom right of Fig.
\ref{fig:data_acess}). Users can express the SEDs in various units 
(e.g., different flux units versus frequency or energy) through the 'Plot Options' 
panel, as shown in Fig. \ref{fig:data_acess}. Additionally, users can apply time 
filtering through the 'Time Filtering' panel (see Fig. \ref{fig:data_acess} right 
side) by providing time constraints, either in the Modified Julian Date 
(MJD) or regular calendar dates, to retrieve data for specific periods. Along with 
the SED, the sub-tabs provide additional data and information about the source 
under investigation. For example, under the 'SED/LC ANIMATION' sub-tab, the 
dynamic changes in the SED composed of simultaneous data (see Section \ref{mseds} 
for details) over time are shown, while the source's position in the sky is 
displayed using the Aladin interactive sky atlas \citep{2000A&AS..143...33B}, 
which is accessible under the 'ALADIN' sub-tab. Additionally, a list of articles 
related to the sources for which data were downloaded is provided under the 
'RELATED ARTICLES' sub-tab, with references retrieved from the SAO/NASA 
Astrophysics Data System\footnote{\url{https://ui.adsabs.harvard.edu}}. Under the 
SED (see Fig. \ref{fig:data_acess}), various information about the source are 
provided, including RA/Dec coordinates, galactic coordinates, and redshift, which 
is retrieved from the NASA/IPAC Extragalactic Database
\footnote{\url{https://ned.ipac.caltech.edu}}.
As  mentioned in Section \ref{blazar}, the peak frequency of the synchrotron emission 
is used to further classify blazars, so  when IR data from NEOWISE are available 
and not contaminated by jet-unrelated emission, the estimated peak frequency of 
the low-energy synchrotron component \citep[W-peak;][]{2024ApJ...963...48G} is
also provided.

The data can be downloaded in CSV format by clicking the 'Download CSV' button. An illustration of the data format is provided in the upper right part of Fig. \ref{fig:data_acess}. The data columns are structured as follows: \textit{freq:} the frequency of the observation, \textit{flux:} the observed flux value, \textit{err\_flux:} the error associated with the flux measurement, \textit{MJD\_start:} the MJD indicating the start of the observation, \textit{MJD\_end:} the MJD indicating the end of the observation, \textit{flag:} indicates whether the flux measurement is an upper limit ('UL'). If the measurement is not an upper limit, this column is left empty, \textit{catalog:} the name of the catalog from which the data is extracted, \textit{reference:} the bibliographic reference associated with the data.

\subsection{Theoretical Modeling}

\begin{figure}
    \centering
    \includegraphics[width=0.98\linewidth]{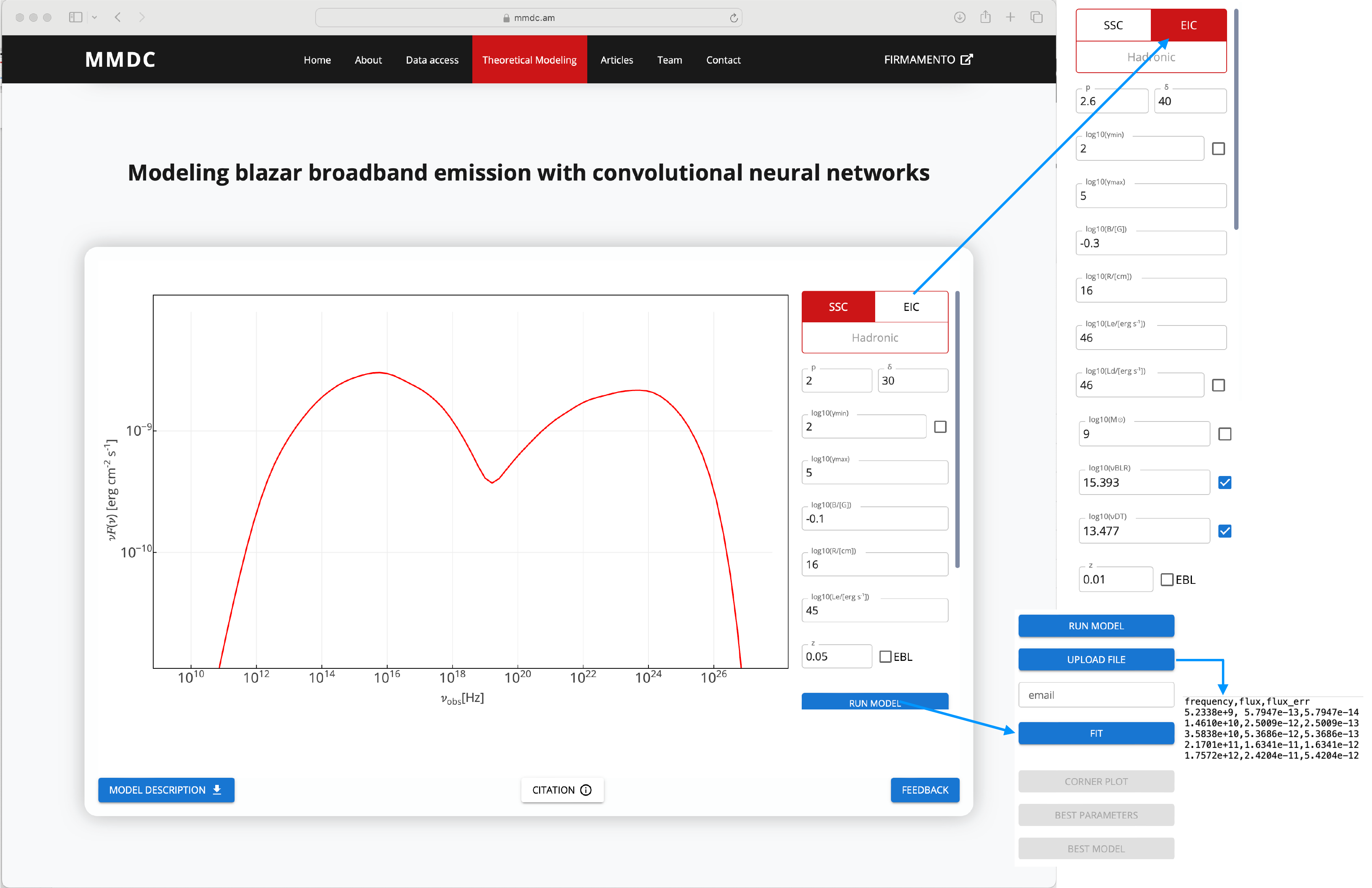}
    \caption{\mmdc\ interface for the Theoretical modeling. The interface allows for the selection of different emission models, such as SSC, EIC, and Hadronic models (available soon), and the generation of models for various input parameters. In this example, the right panels show the input parameters for SSC and EIC. On the right side, the panel for inputting observational data and fitting the model to the data is also shown, along with an example of the data format that can be uploaded.}
    \label{fig:data_modeling}
\end{figure}

Fig. \ref{fig:data_modeling} illustrates the Theoretical Modeling
tab, designed for performing theoretical modeling of blazar broadband
SED. This tab offers two primary functionalities:
(i) generating a SED for a given set of
parameters by completing the necessary input fields, and (ii) fitting
the model to the data provided by the user. The
fitting process, along with the models and corresponding parameters, is
discussed in Section \ref{fig:data_modeling}, with additional details
available in the 'Model Description' field. In the top right corner of
the control panel, the user can select from various theoretical models,
including Synchrotron-Self-Compton (SSC), External Inverse Compton (EIC),
and Hadronic models (the latter available soon). Upon selection, a
corresponding panel with the parameter names appears. For a detailed
description of these parameters, see Section \ref{fig:data_modeling} and
the 'Model Description' field. Examples of panels for SSC and EIC are
shown in Fig. \ref{fig:data_modeling}. After entering the parameters
and clicking the 'RUN MODEL' button, the corresponding SED is displayed.
If the entered parameters fall outside the acceptable range, a red message
appears, indicating the correct range. As an example, the SSC model for
the parameters indicated in the panel is shown in Fig.
\ref{fig:data_modeling}. Users also have the option to specify whether
Extragalactic Background Light (EBL) absorption should be considered. If
selected, the EBL correction is performed using the model proposed by
\citep{2011MNRAS.410.2556D}. The email provided in the email field is
used when the fitting option is selected; the lengthy fitting process runs
in the background, and an email notifies the user once the fit is
complete.
The panel for uploading data is shown at the bottom right, along with the
format specifications for data that can be uploaded for model fitting. The
'CORNER PLOT,' 'BEST PARAMETERS,' and 'BEST MODEL' fields are described in
Section \ref{fig:data_modeling}.

\subsection{Article search}

\mmdc{} incorporates an article search interface (under tab
articles) designed to facilitate the retrieval of relevant scientific
literature based on astronomical objects. This interface provides two
distinct methods for accessing articles: searching by object name and
searching by coordinates in the SAO/NASA Astrophysics Data
System\footnote{\url{https://ui.adsabs.harvard.edu}}. The Object Name
search allows users to perform a search using the precise name of the
astronomical source, which is case- and space-sensitive (e.g., "CTA 102"
and "CTA102" will retrieve different results). This method ensures
that only articles containing the exact source name are retrieved,
making it particularly useful for finding all literature directly
related to a specific object. Conversely, the Coordinates search performs
a cone search around the coordinates of the provided source, with a
default search radius of two arcseconds. This method is effective
for identifying literature that discusses objects within the specified
radius, allowing for the discovery of nearby sources or related
phenomena. Both search functionalities are complemented by
customizable temporal filters, enabling users to constrain their
searches to specific publication years.
 
\subsection{Back End}
The backend architecture of \mmdc{} is meticulously
designed to efficiently manage the complex workflows within the data center,
focusing on data gathering, data storage, and theoretical modeling. The
backend utilizes a fully containerized architecture, leveraging Docker
containers to significantly improve scalability, reproducibility, deployment
flexibility, and long-term sustainability. Each pipeline component operates
within its own isolated container, allowing for seamless integration
into the overall architecture. This approach minimizes the risk of conflicts
and enhances system stability, ensuring consistent runtime environments
across development, testing, and production settings. This fully
containerized backend architecture offers unparalleled deployment
flexibility, enabling containers to be deployed across a wide range of
environments, from local servers to cloud-based platforms, thereby
allowing efficient resource utilization. Moreover, containerization
plays a crucial role in ensuring the long-term viability of the backend
system. By encapsulating dependencies and configurations within
containers, the system can evolve with minimal disruption, accommodating
updates to underlying software, hardware, or computational methods
without requiring a complete redesign.

The main structure and components of the backend are shown in
Fig. \ref{struct}. A PostgreSQL database serves as the central repository
for storing data in \mmdc{}, chosen for its robustness, scalability, and
support for complex queries, making it particularly well-suited for managing
large datasets. The pre-analyzed data in the optical/UV, X-ray, and \gray{}
bands are systematically organized into the PostgreSQL database, which
includes coordinates, frequency, flux and its uncertainty, observation
time (start and end), and can be retrieved by performing
search queries. The data gathering tool executes a container to run
the VOU-Blazar tool, accesses real-time data from ASAS-SN, ZTF, and
Pan-STARRS, and performs searches in the database containing data in
optical/UV, X-ray, and \gray{} bands. The
retrieved data are then stored in the PostgreSQL database
and can be accessed from the frontend. Therefore, if the same source
is searched again and the 'Force Run' option
is not selected, the data will be promptly retrieved from the database
and displayed without executing any processing scripts. Furthermore,
data storage is managed through a global indexing scheme that divides
the entire sky into equal pixels and assigns a unique pixel number.
This process utilizes the ang2pix tool from the HEALPix software
library \citep{2005ApJ...622..759G}, which converts the source
coordinates into pixel indices. Thus, the data are stored with
pixel indexing that correlates with the source coordinates and when
a source with a different name is searched, it is not necessary to
re-run data gathering and the same data will be retrieved.

The backend also uses a containerized tool that employs the
CNN and MultiNest for theoretical modeling. When data is uploaded, it is
transferred to the Docker container, where the data fitting process occurs.
After the fitting is executed, the results are returned to a PostgreSQL
database where the results are stored (see Fig. \ref{struct}).
The fit results are communicated back to the user
through an email which contains a link which reads from the database
and redirects the user to \url{www.mmdc.am}, allowing for visual inspection
of the best-fit parameters and retrieval of the best model. The
parameter posterior distributions are also available for download.

\subsection{The server}
The website of \mmdc{} (\url{www.mmdc.am}) is hosted within
the Academic Scientific Research Computer Network of Armenia (ASNET;
\url{www.asnet.am}) which serves as a critical component
of our system. When a user initiates a new job in the frontend—whether
requesting data or performing modeling—the request is received by this
server, which validates and redirects it for further processing. This
server also stores the generated SED/LC animations, indexed in the same
manner as the SEDs. Therefore, when a user searches for data, if the
corresponding animation is available under the same index, it is served
to the frontend. Furthermore, if the requested job has not been
processed (e.g., if the requested SEDs are not available, 'Force Run'
is selected, or theoretical fitting is requested), the server sends a
request to a compute node to execute the
corresponding process. Once the job is completed,
the results are returned to this server and served to the frontend.
The computations (compute node) are executed using computational resources allocated within ASNET (subject to change; a separate high-performance server will be allocated for the computations). Currently, up to 3 jobs (fitting) can run simultaneously. However, to manage the workload effectively, a queue system is implemented, allowing requests to be stored and processed sequentially as server availability permits.
\section{\mmdc: Multiwavelength and multi-temporal SEDs}\label{mseds}

The data acquisition processes and analyses
described in the previous sections offer a unique opportunity to retrieve
broadband data from blazar observations contained in the catalogs, as well
as multitemporal data newly analyzed in the optical/UV,
X-ray, and \(\gamma\)-ray bands. This provides a
comprehensive display of blazar emission components as well as a detailed view of the dynamical changes in the radiation component over time.

\begin{figure*}
     \centering
    \includegraphics[width=0.45\textwidth]{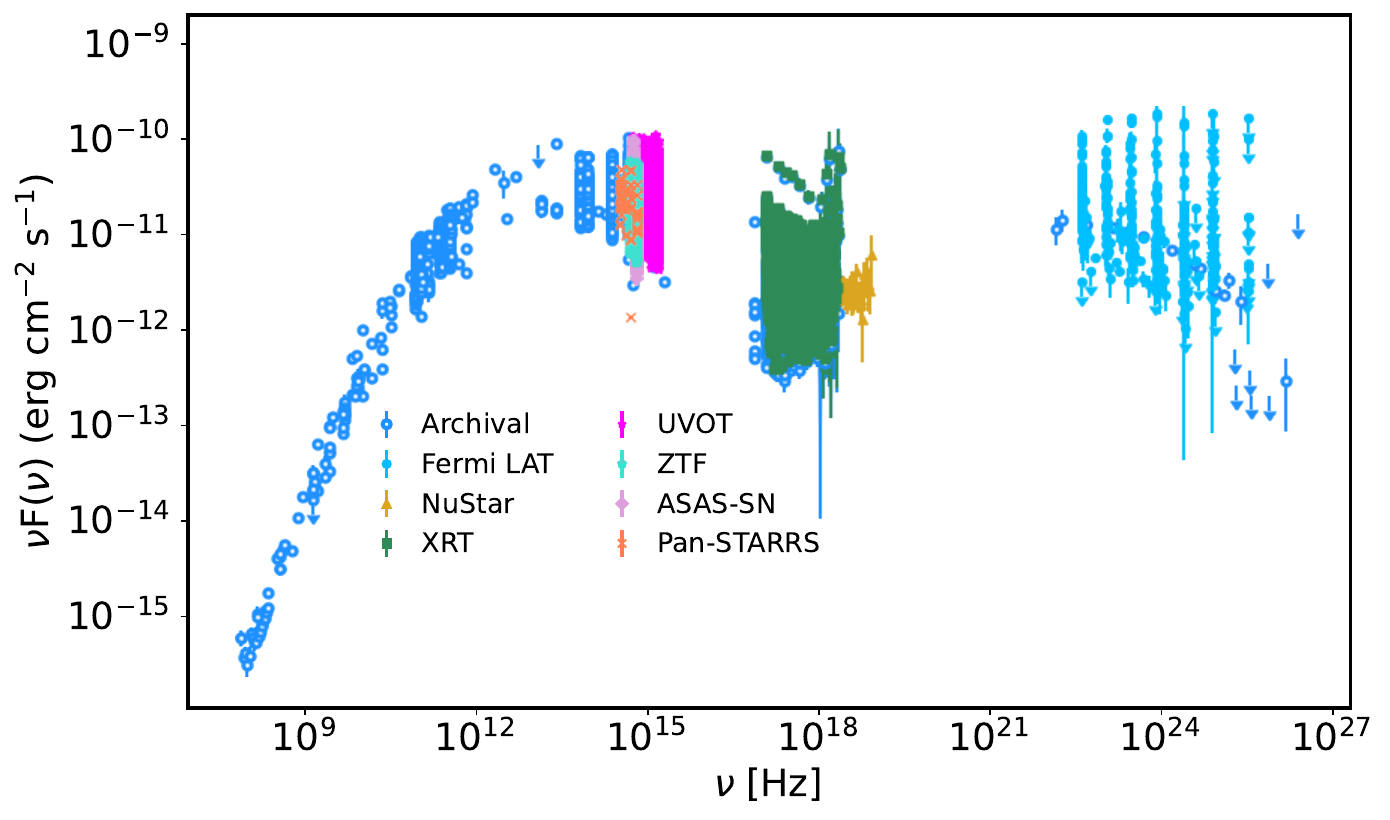}
     \includegraphics[width=0.45\textwidth]{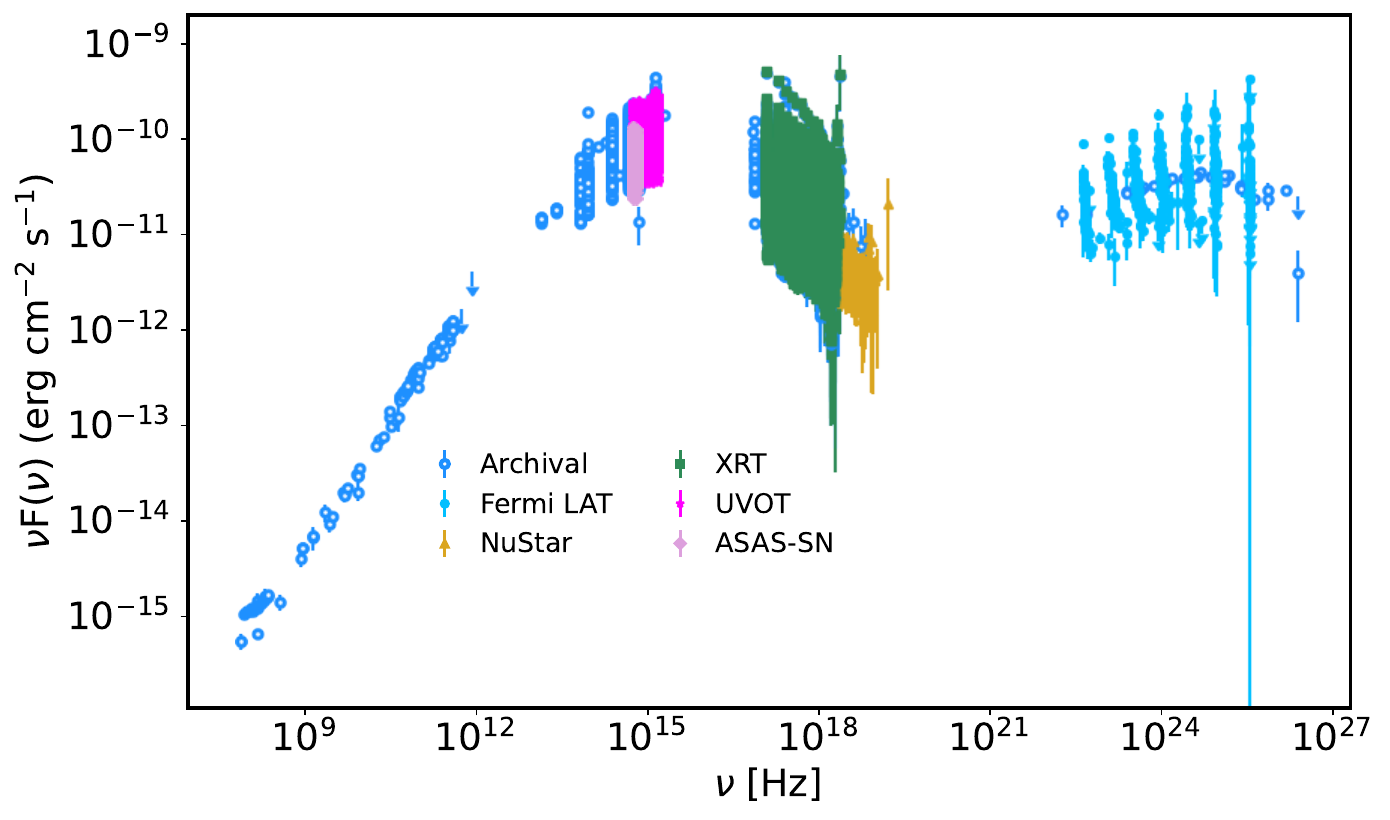}
     \caption{The broadband SEDs of OJ 287 and PKS 2155-304 are displayed using data retrieved from \mmdc. The blue data points represent archival data, while the other data available from \mmdc\ are depicted in various colors, as shown in the legend. Note that the scaling of both figure is the same to ease direct comparison.}\label{MW_SED}
\end{figure*}

In Fig. \ref{MW_SED}, the broadband SEDs of OJ 287 and PKS 2155-304 are
displayed, using data from \mmdc. The newly analysed data
in the \(\gamma\)-ray, X-ray, and optical/UV bands available from \mmdc\ are
shown in various colors: \(\gamma\)-ray data from \textit{Fermi}-LAT in sky blue,
NuStar data in goldenrod, XRT data in green, UVOT data in magenta, ZTF data
in turquoise, and ASAS-SN data in plum. For both sources, the
provided data enables the constraints
on emission components and highlight the large flux and spectral variability. For instance, for
OJ 287, the emission in the optical/UV varies by two orders of magnitude, while
in the X-ray and \(\gamma\)-ray bands, it varies by nearly three orders
of magnitude. A similar variability pattern is observed for PKS 2155-304.
This shows the richness of the data provided by \mmdc\ and how they can
be used to study the properties of different blazars.

The data presented in Fig. \ref{MW_SED} presents a general view of the emission components
and their potential amplitude variations. However, for a
comprehensive understanding of the nature of blazar
emissions, a detailed investigation of the
dynamical changes in the various bands
is necessary. This involves examining what physical changes lead to
variations in emissions across different bands. To facilitate an in-depth
study of blazars, the time evolution of emission components (SED/Light
curve animations) is also made available alongside the SEDs. This involves
binning the observed data, particularly when the observational times
are known, in small intervals to construct quasi-simultaneous SEDs of
the same object. The periods to generate SEDs are selected based on the
Bayesian intervals of the \(\gamma\)-ray light curves
to ensure the spectrum extends to HE \(\gamma\)-ray band. For each
Bayesian interval, all available data in other bands are considered,
which of course cover shorter periods than the total Bayesian interval.
Then, an 'adaptive' scan is performed within the Bayesian interval; that is,
a small time bin (\(\Delta t\)) is added to the beginning of the interval.
Subsequently, the condition that there are
observed data in at least three frequency intervals (i.e.,
considering the frequencies in  radio, IR,
optical, X-ray and \gray\ bands) is verified. If it is not satisfied, the time
is incremented by \(\Delta t\) until the
condition is satisfied. Thus, if there are many measurements within the
Bayesian interval, they will not be grouped into one; instead, they will
be considered and displayed sequentially.

The time increment, \(\Delta t\), is defined in two different ways,
depending on the availability of \textit{Swift} observations. If \textit{Swift} observations
are present in the Bayesian block interval, \(\Delta t\) is set to
half of the minimum difference between consecutive \textit{Swift} observations. This
ensures that all \textit{Swift} observations will be displayed separately. When there
are no \textit{Swift} observations, all observations are first ordered in decreasing
temporal order. Then, the difference between successive observation times,
\(t_{\rm n+1}\) and \(t_{\rm n}\), is calculated. The mean of these differences
is computed and subsequently assigned to \(\Delta t\). Then, by displaying the
data together and changing it over time, we can visualize the evolution of
the spectral components. In the past, such SED/Light
curve animations were already generated for 3C 454.3, BL Lacertae, and CTA 102,
and were used to investigate the dynamical changes in their emission
components, providing a quantitative approach to studying
the emissions from these blazars \citep{2021MNRAS.504.5074S, 2022MNRAS.517.2757S,
2022MNRAS.513.4645S}. These animations are accessible through \mmdc\, under
the panel data access SED/LC animation, and some of them are available by following links : \href{https://www.youtube.com/watch?v=wMZRMuaScwk}{\nolinkurl{Mkn 421}}, \href{https://www.youtube.com/watch?v=HGdfIlzmBtQ}{\nolinkurl{Mkn 501}}, \href{https://www.youtube.com/watch?v=wB7Nbyam5nQ}{\nolinkurl{PKS 1510-089}} , and \href{https://www.youtube.com/watch?v=7RYfWqFsmws}{\nolinkurl{3C 273}}.

\section{MMDC: blazar SED modeling}
\label{modeling}

The extensive data from blazar observations provide new perspectives for
theoretical interpretation, allowing for the modeling of multiple periods
and making it possible to retrieve the dynamic changes in the emission
components and parameters. Such approach allows
to investigate the processes through which particles
interact and lose energy, as well as study their evolution
in time, thereby enhancing our understanding of the acceleration and
cooling processes occurring in blazar jets. However, the models used to
explain the observed data have become more complex, as they now include
various processes to account for the observed features. This complexity, has made the models computationally intensive, limiting
the exploration of parameter space and direct data fitting. Direct fitting
becomes feasible when an ad-hoc assumption about the emitting particles
is made. To overcome the challenges associated with data modeling using
self-consistent models, \citet{2024ApJ...963...71B} and \citet{2024arXiv240207495S} introduced a novel method for modeling
the SEDs of blazars \citep[see also][]{2024A&A...683A.185T}.
This method employs CNNs trained on leptonic
models that incorporates both synchrotron and inverse
Compton emission processes, considering both
internal and external photon fields, as well as accounting for
self-consistent electron cooling and pair creation–annihilation processes.
The CNN is capable of reproducing the radiative signatures of particle
emissions with a high accuracy while significantly reducing the computational time.
This efficiency enables the fitting of multiwavelength data sets within
reasonable time scales. This methodology significantly differs from other
online tools, as
it does not rely on ad-hoc assumptions about the electron spectrum. Instead, the spectrum is computed based on particle
injection and cooling.

Our machine-leaning-based models are accessible
to the community through \mmdc, allowing users to upload and fit their data.
The web interface currently enables the selection between the
Synchrotron Self-Compton (SSC) model, External Inverse Compton (EIC) model,
and soon Lepto-hadronic models. A brief
description of each of the currently available models
is provided below, with detailed information available in \citet{2024ApJ...963...71B} and \citet{2024arXiv240207495S}.
\begin{itemize}
  \item {\it SSC:} The synchrotron self-Compton model is
  frequently employed to model the broadband emission observed in BL Lacs. In
  this framework, the low-energy component of the emission spectrum is
  interpreted as synchrotron radiation from electrons, while the HE
  component arises from inverse Compton scattering of synchrotron photons
  by the same population of electrons \citep{1996ApJ...461..657B,
  1985A&A...146..204G, 1985ApJ...298..114M, 1999MNRAS.306..551C}. The emission 
  is assumed to originate from a spherical region filled by a homogeneous and constant
  magnetic field, which moves relativistically with
  a Lorentz factor, \(\Gamma\) (for a small viewing angle, the
  Doppler boost is such that \(\Gamma\simeq\delta\)). We assume that
  electrons are
  injected into this emitting region with a cutoff power-law energy
  distribution. They then interact and cool. The
  temporal evolution of the electron distribution is governed by a
  Fokker–Planck diffusion equation, whereas the photon evolution is
  described by an integro-differential equation. These equations are solved
  using {\it SOPRANO} \citep[Simulator Of Processes in Relativistic AstroNomical Objects,][]{2022MNRAS.509.2102G}, with the system being evolved
  over time until \(t = 4 t_{\rm dyn} = 4R/(\delta c)\) in order to represent a steady state.
  
  Accordingly, this model comprises
  seven free parameters: the comoving blob radius (\(R\)), the Doppler factor
  (\(\delta\)) of the emission region, the comoving magnetic field strength
  (\(B\)) within the emission zone, the electron luminosity (\(L_e\)),
  the minimum Lorentz factor (\(\gamma_{\text{min}}\)), the cutoff Lorentz
  factor (\(\gamma_{\text{max}}\)), and the power-law index (\(p\)). For each
  parameter, a wide range that is relevant to the SSC model is considered \citep[refer
  to Table 1 in][]{2024ApJ...963...71B}, and the corresponding outputs
  (SEDs) for a set of parameters are computed using {\it SOPRANO}. Our training set consists of \(2 \times 10^5\) such spectra \citep[for details
  on the network structure and training process, see][]{2024ApJ...963...71B}.
  The CNN effectively learns the relationship between input parameters and
  their corresponding spectra, and can be used for SED modeling significantly
  faster than when directly using {\it SOPRANO} or any similar tool. Subsequently,
  the trained CNN is used in conjunction with MultiNest \citep{2008MNRAS.384..449F, 2009MNRAS.398.1601F, 2019OJAp....2E..10F} for data fitting, enabling the determination of the
  best-fit parameters that explain the observed data, as well as enabling the analysis of the parameter posterior distributions.

  An example of fitting the broadband SED of Mrk 501 using \mmdc\ is illustrated in Fig. \ref{SED_SSC}. The uploaded data are displayed in blue, the best-fit model in red, and the uncertainties in grey. The best-fit parameters are shown in the left panel and can be downloaded as a CSV file (under the "BEST PARAMETER" button). Additionally, the best model can be downloaded as a CSV file (under the "BEST MODEL" button), while parameter posterior distributions are available under the "CORNER PLOT" button. The plot is interactive, allowing users to zoom in and closely examine how the model accounts for the data within a specific interval.

\begin{figure*}
     \centering
    \includegraphics[width=0.98\textwidth]{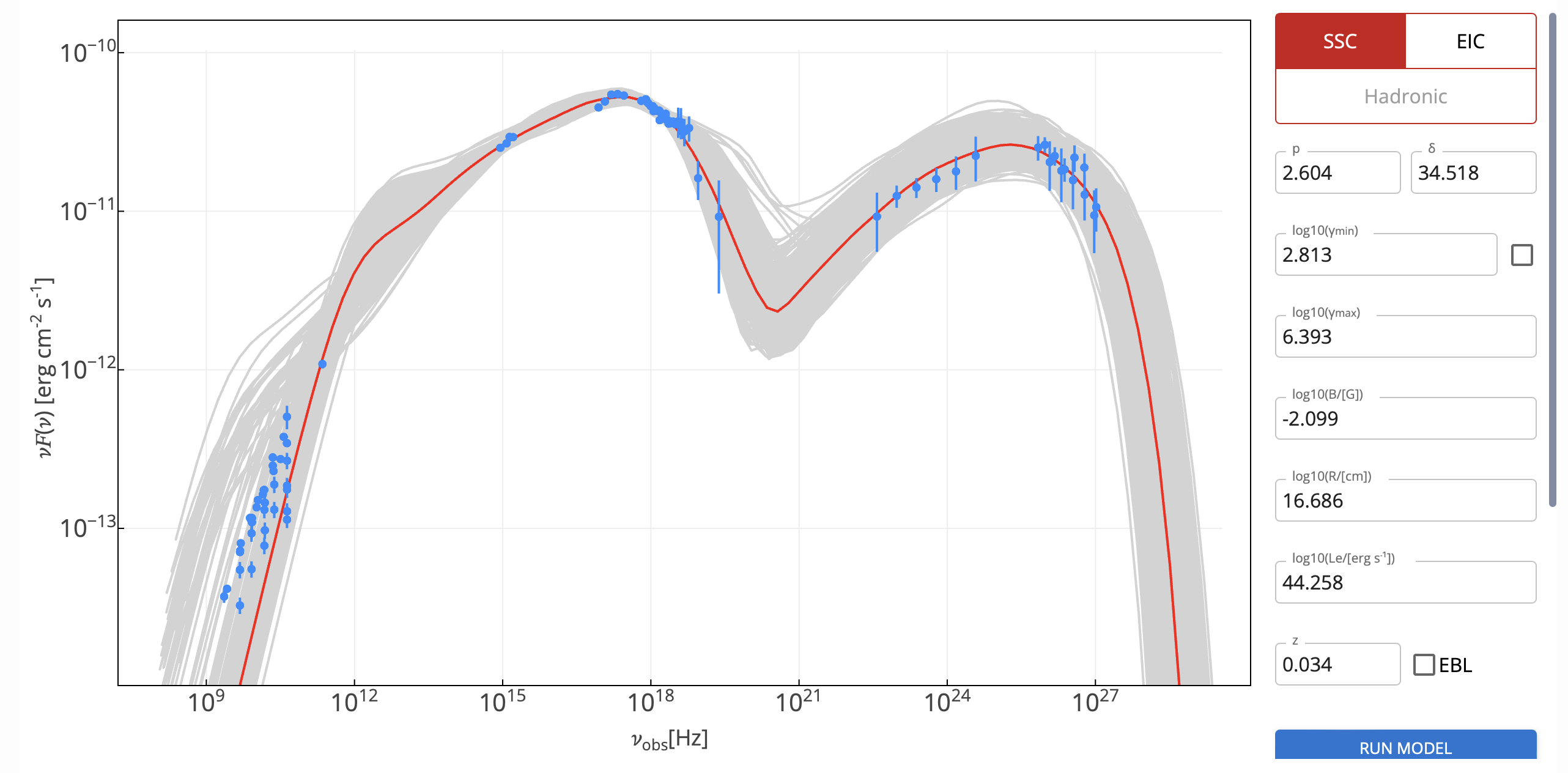}
     \caption{The SED of Mrk 501 modeled within a one-zone leptonic scenario using \mmdc. The red line represents the model with the best-fit parameters, that is, when the likelihood is maximized, and the gray spectra depict the model's uncertainty. The uploaded data are shown in blue.} \label{SED_SSC}
\end{figure*}

  \item {\it EIC:}  The External Inverse Compton (EIC) model is usually used to model the SED of FSRQs \citep{blazejowski,ghiselini09, sikora}. For FSRQs, external photon fields can
  be dominant regarding cooling and the
  formation of the HE component. These photons may originate directly
  from the accretion disk, be reflected from the broad line region (BLR),
  or be emitted directly from the dusty torus
  \citep[e.g.,][]{blazejowski,ghiselini09, sikora}. Considering the contribution
  of these external photons, the cooling rate of electrons is altered, which
  also affects the resultant radiative signature. Compared to
  the SSC scenario,
  this model introduces additional free parameters (7 versus 11), complicating
  both the initial simulations and the learning process of the CNN. The
  methodology, namely, the simulation of particle radiative signatures across
  a wide range of parameters with SOPRANO, remains the same as with the SSC
  case, albeit with the addition of the parametrization of external fields, for details see \citet{2024arXiv240207495S}. The EIC model incorporates eleven free parameters. In
  addition to the parameters of the SSC model, it further includes
  parameters such as the accretion disk luminosity ($L_d$), the mass of
  the central supermassive black hole ($M_{\mathrm{BH}}$), and the
  temperature/frequency characteristics of the broad-line and dusty torus
  regions emission. Considering both the large number
  of parameters and their  broad range \citep[see Table
  1 of][]{2024arXiv240207495S}, $10^6$ spectra have been generated using
  {\it SOPRANO} and employed to train the CNN. Given that
  the spectra appears more scattered this time, to
  achieve adequate accuracy, the layers and dimensions of the CNN were
  slightly modified in comparison to those used for the SSC model. The
  developed CNN can accurately reproduce the radiative signatures of
  electrons when the emitting region is at various distances from the
  central source, thus allowing for different photon fields
  to contribute. The CNN
  can therefore be used to model the SED of FSRQs.
  
  Similar to the SSC model, this new CNN is accessible via \mmdc, enabling users
  to fit their uploaded data. The interface and methodology of implementation
  mirrors that of the SSC model, with the addition of extra parameters available
  in the panel. The spectral modeling of 3C 279
  during the flaring period
  is shown in Fig. \ref{SED_EIC}. During the fitting process, the frequencies
  of the emissions from the broad-line and dusty torus regions were fixed at
  $2.47\times10^{15}$ Hz and $3.0\times10^{13}$ Hz, respectively, along with
  $M_{\mathrm{BH}}=7.9\times10^8\: M_{\odot}$. The optical/UV and X-ray data
  are interpreted as synchrotron/SSC emissions, while the HE \gray\ data are
  interpreted as resulting from external inverse Compton scattering. All the
  fitting results described for the SSC case that are available for download are also accessible in this case.

  \begin{figure*}
     \centering
    \includegraphics[width=0.98\textwidth]{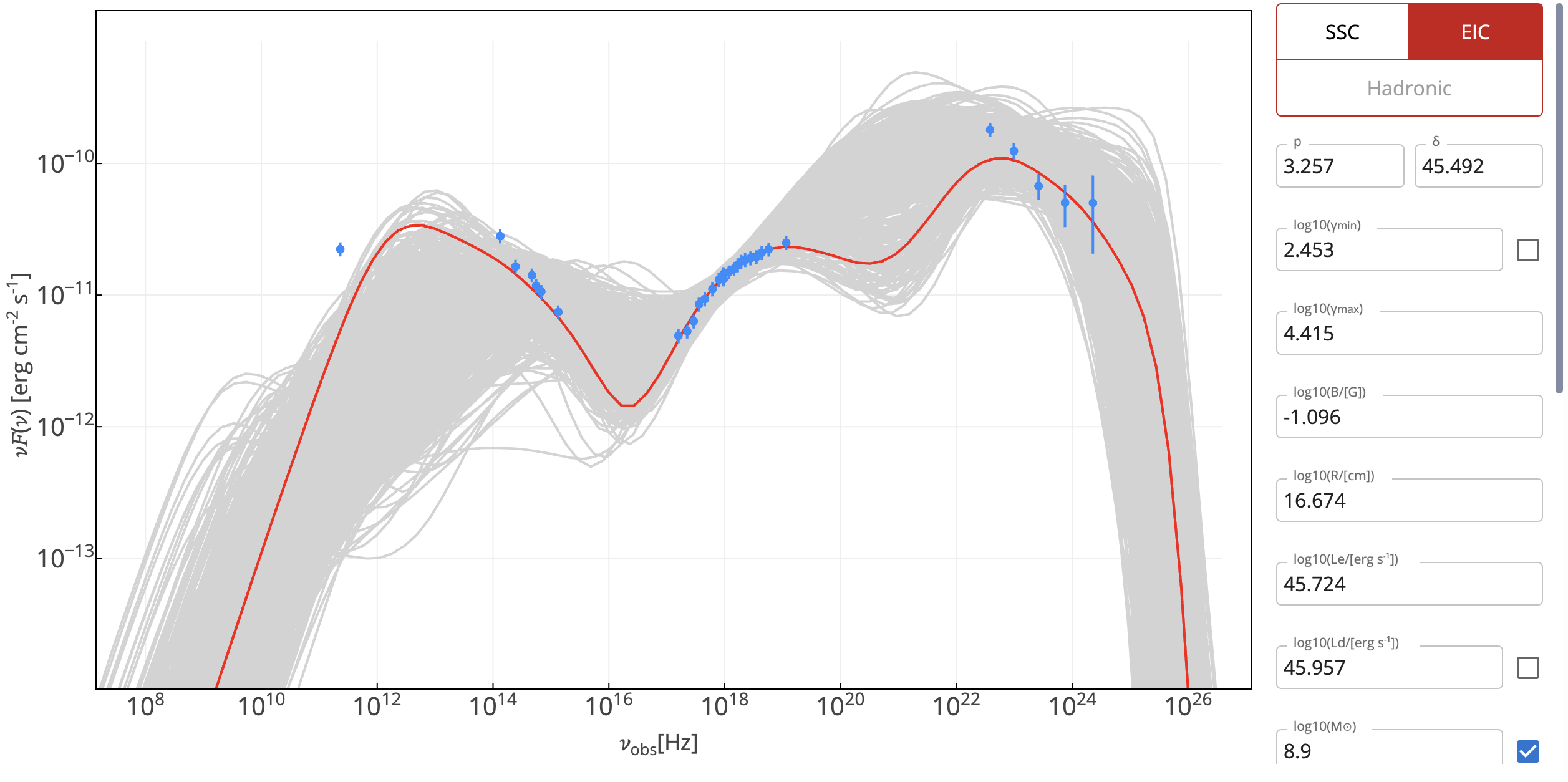}
     \caption{The SED of 3C 279 during the flaring period, modeled within a one-zone leptonic EIC scenario using \mmdc. The color coding is the same as in Fig. \ref{SED_SSC}.}\label{SED_EIC}
\end{figure*}

  \item {\it Lepto-hadronic models:} In the hadronic or lepto-hadronic models
  of blazar emission, the HE component is produced by the contribution of
  protons, either through direct synchrotron emission \citep{2001APh....15..121M}
  or via the secondaries generated in photo-pion and photo-pair interactions
  \citep{1993A&A...269...67M, 1989A&A...221..211M, 2001APh....15..121M, mucke2,
  2013ApJ...768...54B, 2015MNRAS.447...36P, 2022MNRAS.509.2102G}. Following the
  observations of VHE neutrinos from the direction of TXS
  0506+056 \citep{2018Sci...361..147I, 2018Sci...361.1378I, 2018MNRAS.480..192P}
  and PKS 0735+178 \citep{2023MNRAS.519.1396S, 2023ApJ...954...70A}, hadronic, and especially
  lepto-hadronic, models have gained prominence since
  they also predict VHE neutrino emission. We are planning to shortly include hadronic models to \mmdc, allowing users to fit multi-messenger dataset comprising the photon SEDs and estimate of neutrino flux.
  \end{itemize}
\section{Conclusions and Future Perspectives}
\label{conc}

The continuous increase of astrophysical data creates excellent conditions for research, allowing us to broaden our understanding of the Universe. However, considering that usually only raw data is accessible, its efficient and timely analysis, due to its size and complexity, introduces challenges for the effective usability of the observed data. To enhance and facilitate blazar research, \mmdc\ provides a large volume of science-ready multiwavelength data observed over different periods, which can be visualized interactively to help users easily grasp the complex properties of the data. This data, available for a large number of sources, can be used to retrieve emission features in different bands and for theoretical modeling, enabling more efficient and advanced research on blazars.

\mmdc\ contains extensive data from blazar multiwavelength observations. Archival data from blazar observations are retrieved using the VOU-Blazars tool, while data in the optical/UV (from \textit{Swift}-UVOT), X-ray (\textit{Swift}-XRT, NuSTAR), and \gray\ band (\textit{Fermi}-LAT) are thoroughly analyzed and made available, with additional data accessed in real-time from ASAS-SN, ZTF, and Pan-STARRS. For example, \textit{Fermi}-LAT data analyzed for various periods between 2008 and 2023 provides the most detailed view of the source \gray\ emission spectral changes in various flaring and quiescent states. Moreover, \mmdc\ allows the construction of time-resolved SEDs, enabling researchers to examine the emission characteristics of blazars across different periods. This temporal resolution provides insights into the dynamic changes in blazar emissions and helps in understanding the underlying physical processes.

Additionally, \mmdc\ offers a unique feature for modeling SEDs with self-consistent models using CNNs trained on different theoretical models, including SSC, EIC, and soon Lepto-hadronic models. This advanced modeling capability allows for efficient and accurate fitting of multiwavelength and multimessenger data and retrieving model free parameters, enabling in-depth analysis of blazar emissions. By integrating these sophisticated modeling tools, \mmdc\ effectively combines the accessibility of a large amount of data with the ability to interpret observational data, leading to new discoveries and a deeper understanding of blazar physics.

To expand the potential for multiwavelength and multimessenger studies
of blazars, \mmdc{} will soon interface with an artificial
intelligence tool. Rapid advancements in Large Language Models (LLMs)
have recently demonstrated remarkable capabilities in natural language
processing tasks, significantly contributing to various fields, including
science, and driving innovation. In this context, \texttt{astroLLM} is
currently under development with the primary goal of creating a powerful
research assistant tailored specifically to blazar studies. \textit{astroLLM}
is designed using retrieval-augmented generation (RAG) techniques, which
combine pre-trained language models with the dynamic retrieval of
relevant astrophysical literature and datasets on blazars. This approach
enables the application of LLMs trained on a vast number of parameters
to specific domain challenges, effectively overcoming difficulties related
to technical terminology and concepts. In addition to generating human-like
text, \texttt{astroLLM} will be capable of retrieving multiwavelength and
multimessenger data from \mmdc{} and performing high-level theoretical
modeling through AI agents, significantly advancing blazar research.
Moreover, \texttt{astroLLM} will serve as an educational tool, offering
powerful capabilities for training and knowledge dissemination.
Initially, \texttt{astroLLM} will focus solely on blazar science, with
its functionality gradually expanding to include data retrieval and
modeling capabilities for other classes of astrophysical sources.


\begin{acknowledgments}
We thank the referee for the comments and suggestions that helped to improve the paper and \mmdc{} tool. The research was supported by the Higher Education and Science Committee of MESCS RA (Research project No 23LCG-1C004). DB and HDB acknowledge support from the European Research Council via the
ERC consolidating grant $\sharp$773062 (acronym O.M.J.).
\end{acknowledgments}

%

\vspace{5mm}
\facilities{Fermi-LAT, NuSTAR, Swift(XRT and UVOT), ASAS-SN, ZTF, 
Pan-STARRS}

\software{FermiPy \citep{2017ICRC...35..824W}, VOU-Blazars \citep{2020A&C....3000350C}, Aladin \citep{2000A&AS..143...33B}, Sky Patrol V2.0 \citep{2023arXiv230403791H}, swift\_xrtproc \citep{2021MNRAS.507.5690G}, NuSTAR\_Spectra pipeline \citep{2022MNRAS.514.3179M}
 }
 



\appendix

\section{Automated \textit{Fermi}-LAT Data Analysis}\label{appendix}
In the current study, PASS8 \textit{Fermi}-LAT data collected between August 4, 2008,
and July 4, 2023 (MET 239667417-710178221) were downloaded for
all selected sources. The data
were filtered and analyzed using the standard binned likelihood analysis method\footnote{For a description of this method, see \url{https://fermi.gsfc.nasa.gov/ssc/data/analysis/documentation/}}.
The analysis for each source was performed separately, in
the energy range from 100 MeV to 300 GeV considering
the events from a region of interest (ROI) of $12^\circ$ centered on the
\gray\ position of each source. The ROI was visually checked during the
analysis, and in some cases, it was reduced to $10^\circ$ to better represent
the region and the distribution of sources within it. The analysis was conducted using the
Fermi ScienceTools version 2.0.8 and the P8R3\_SOURCE\_V3 instrument
response function. Only events with a high probability
of being photons were selected, using the cut {\it evclass = 128} and
{\it evtype = 3}, and the filter ${\rm (DATA\_QUAL>0)\&\&(LAT\_CONFIG==1)}$
was applied to update the good time interval based on spacecraft specifications.
Further, to reduce the \grays\ from the Earth's limb, a maximum zenith angle
cut of $>90^\circ$ was also applied. To build a model that describes the
ROIs, the Fermi fourth source catalog (4FGL) incremental version
\citep[DR 3;][]{2022ApJS..260...53A} was used. All the sources within
the ROI around each blazar, plus an annulus of $5^\circ$, were included
in the model file. To determine the spectral characteristics
of the source, the spectral parameters of all sources in
the ROI (i.e., within a $12^\circ$ radius) were left free, while
those of the sources outside the ROI ($12^\circ-17^\circ$) were fixed. The model also includes the galactic
background and isotropic galactic emissions, which were modeled with
the latest available versions of the files, gll\_iem\_v07 and
iso\_ P8R3\_SOURCE\_V3\_v1, respectively. The resultant spectral
model was fitted to the data using the binned likelihood analysis method
and the {\it fermiPy} tool \citep{2017ICRC...35..824W}. In the all-time 
analysis, the spectra of the sources under
investigation were modeled using the same model as in the
4FGL. However, when the catalog spectrum is not a power-law,  an additional analysis was conducted, assuming the spectrum could be approximated by a power-law. This latter model was utilized in the light-curve calculations, as it provides a good approximation of the spectrum when data analysis is applied to shorter periods.

Next, the variability of each source in the \gray\ band
was investigated by binning the light-curves using an
adaptive binning method. Indeed,
a traditional fixed-time binning method tends
to smooth out rapid variations due to long bins, while short bins might result
in many upper limits during periods of low activity. In contrast, the
adaptive binning method allows for the duration of each bin to be
flexibly adjusted, ensuring that bins have a constant flux uncertainty
above the optimal energies. Therefore, when the source is in a bright
emission state, the bins are shorter, whereas longer bins are used during
lower and/or average source states. Light curves generated through this
method have been extensively used to study the variability of blazars,
demonstrating the capability to identify short-timescale flux variations \citep[see, e.g.,][]{2013A&A...557A..71R, 2016ApJ...830..162B, 2017MNRAS.470.2861S, 2017A&A...608A..37Z, 2017ApJ...848..111B, 2018ApJ...863..114G, 2018A&A...614A...6S, 2021MNRAS.504.5074S, 2022MNRAS.517.2757S, 2022MNRAS.513.4645S}.

The continuous observation of all sources with the \textit{Fermi}-LAT instrument
presents an unprecedented opportunity to investigate \gray\ flux changes
over different periods. The light curves provides
information on flux and photon indices over short periods, but conducting a
detailed spectral analysis for these periods is challenging: some bins are
too short, resulting in only a few spectral points, each with
significant uncertainty. Motivated by the necessity to have a detailed
view of the spectral changes in the \gray\ band, the light curves produced
with the help of the adaptive binning method were divided into piece-wise
constant blocks \citep[Bayesian blocks, ][]{2013ApJ...764..167S}. Each
interval represents a period during which the flux remains constant. This
approach allows for the merging of periods when the source emission does
not change significantly, thereby constructing longer periods with
consistent flux. The spectrum of each source
is computed within each Bayesian interval by applying an unbinned
likelihood analysis and executing {\it gtlike} separately for either 5
or 7 energy bins. The selection between 5 or 7 bins depends on the
source significance, as measured by the test statistic (TS), defined by \(TS = 2(\ln L_1 - \ln L_0)\), where
\(L_1\) and \(L_0\) represent the maximum likelihoods with and without
the source, respectively \citep{1996ApJ...461..396M}. When TS is
less than 100, 5 energy bins are considered; otherwise, 7 energy
bins are used.


\bibliography{biblio}{}
\bibliographystyle{aasjournal}



\end{document}